\documentclass[lettersize,journal]{IEEEtran}
\usepackage{cite}
\usepackage{amsmath,amssymb,amsfonts}
\usepackage{graphicx,color}
\usepackage{textcomp}
\usepackage{url}
\usepackage{orcidlink} 
\usepackage{algorithm}
\usepackage{algcompatible}
\usepackage{subfigure}

\begin{document}

\title{Streamlined Swift Allocation Strategies for Radio Stripe Networks}

\author{Filipe Conceição\orcidlink{0000-0001-7994-6090}\IEEEauthorrefmark{1}\IEEEauthorrefmark{2},~\IEEEmembership{Graduate Student Member,~IEEE}, Marco Gomes\orcidlink{0000-0003-1124-525X}\IEEEauthorrefmark{1}\IEEEauthorrefmark{2},~\IEEEmembership{Senior Member,~IEEE}, Vitor Silva\orcidlink{0000-0003-2439-1184}\IEEEauthorrefmark{1}\IEEEauthorrefmark{2}, Rui Dinis\orcidlink{0000-0002-8520-7267}\IEEEauthorrefmark{1}\IEEEauthorrefmark{3},~\IEEEmembership{Senior Member,~IEEE}

\IEEEauthorblockA{\IEEEauthorrefmark{1}Instituto de Telecomunica\c{c}\~{o}es, Portugal\\}
\IEEEauthorblockA{\IEEEauthorrefmark{2}Department of Electrical and Computer Engineering, University of Coimbra, \mbox{3030-290} Coimbra, Portugal\\} 
\IEEEauthorblockA{\IEEEauthorrefmark{3}Faculdade de Ciências e Tecnologia, Universidade Nova de Lisboa, \mbox{2829-516} Caparica, Portugal\\} 

\thanks{FCT/MEC funds this work through national funds and when applicable co-funded by European Regional Development Fund (FEDER), the Competitiveness and Internationalization Operational Programme (COMPETE 2020) of the Portugal 2020 framework, Regional OP Centro (POCI-01-0145-FEDER-030588) and Regional Operational Program of Lisbon (Lisboa-01-0145-FEDER-030588) and Financial Support National Public (OE) through national funds, Fundação para a Ciência e Tecnologia, I.P, under the projects 10.54499/UIDB/50008/2020 (DOI identifier: \url{https://doi.org/10.54499/UIDB/50008/2020}), UIDP/50008/2020 and the PhD grant 2020.08124.BD.}
}

\maketitle

\begin{abstract}
This paper proposes the use of an access point (AP) selection scheme to improve the total uplink (UL) spectral efficiency (SE) of a radio stripe (RS) network. This scheme optimizes the allocation matrix between the total number of APs' antennas and users' equipment (UEs) while considering two state-of-the-art and two newly proposed equalization approaches: centralized maximum ratio combining (CMRC), centralized optimal sequence linear processing (COSLP), sequential MRC (SMRC), and parallel MRC (PMRC). The optimization problem is solved through a low-complexity and adaptive genetic algorithm (GA) which aims to output an efficient solution for the AP-UE association matrix. We evaluate the proposed schemes in several network scenarios in terms of SE performance, convergence speed, computational complexity, and fronthaul signalling capacity requirements. The COSLP exhibits the best SE performance at the expense of high computational complexity and fronthaul signalling. The SMRC and PMRC are efficient solutions alternatives to the CMRC, improving its computational complexity and convergence speed. Additionally, we assess the adaptability of the MRC schemes for two different instances of network change: when a new randomly located UE must connect to the RS network and when a random UE is removed from it. We have found that in some cases, by reusing the allocation matrix from the original instance as an initial solution, the SMRC and/or the PMRC can significantly boost the optimization performance of the GA-based AP selection scheme.
\end{abstract}

\begin{IEEEkeywords}
Access Point (AP) Selection, Genetic Algorithm (GA), Maximum Ratio Combining (MRC), Parallel Allocation, Radio Stripe (RS), Resource Allocation (RA), Sequential Allocation, Spectral Efficiency (SE).
\end{IEEEkeywords}

\section{INTRODUCTION}
\IEEEPARstart{M}{assive} multiple-input-multiple-output (mMIMO) is crucial in fifth-generation (5G) New Radio released by the 3rd Generation Partnership Project (3GPP), due to its spatial multiplexing capabilities that allow a considerable increase in spectral efficiency (SE) and power efficiency (PE) \cite{MIMO_Bjornson, MIMO_Larsson, MIMO_Bjornson2}. It consists of the deployment of a massive number of transmitter antennas in a single base station (BS) serving a large number of user equipment (UE) devices within a cell, in the same time-frequency resources. Due to this conventional cell-centric design, a new UE must adapt to the network configuration, and the quality of service they receive largely depends on their position within the network. Hence, inter-cell interference from this deployment leads to poor signal quality at the cell edges, creating a bottleneck for improving the signal-to-interference-and-noise ratio (SINR) during data transmission and reception with UEs, which is primarily linked to their location within the network coverage area.

To provide UEs a ubiquitous experience, future beyond-5G (B5G) network technologies should focus on user-centric mMIMO designs \cite{6G_You, 6G_Saad, 6G_Letaief, 6G_Zhang, 6G_Giordani, LIS_Hu}. In that context, the concept of distributed mMIMO (D-mMIMO) was envisioned, whose main concept is to distribute the BSs' antennas in a large geographical area and allow them to efficiently cooperate in order to serve the UEs \cite{DM_Madhow, DM_Ammar}.

The cell-free (CF) mMIMO network topology was proposed to materialize the D-mMIMO concept \cite{CFvsSC_Ngo, MMSE_Bjornson, CF_Chowdhury, CF_Nayebi, CF_Buzzi, CF_Ngo, CF_Bashar, CF_Bjornson, MH_Conceicao}. The distributed nature of the CF network requires coherent cooperation between the access points (APs) antennas, which can be achieved by considering either a single central processing unit (CPU) \cite{MMSE_Bjornson} or multiple CPUs. However, the CF architecture relies on an impractical assumption regarding the fronthaul link capacity. With each new AP added to the network, an independent fronthaul link must be connected to the CPU. Furthermore, as the number of UEs increases in the network, the load on each fronthaul link escalates. This leads to a non-scalable D-mMIMO network especially in ultra-dense scenarios \cite{RS_Interdonato}. 

One way to circumvent this problem is to implement a distributed scheme, wherein a single fronthaul link is required to connect all network elements, facilitating the sharing of data and control signals between the APs and the CPU. This concept is the basis of the radio stripe (RS) technology \cite{RS_Interdonato, RS_Chiotis, RS_Lakkimsetti, RS_Yuan, RS_Martins, RS_Conceicao, Bi_Conceicao, RS_Mishra, RS_Shaik, RS_Lopez, RS_Ma, RS_Miretti, RS_Shaik2, APS_Conceicao}, which has been recently proposed, being regarded as a cost-effective evolution of D-mMIMO. In an RS, the APs are strategically positioned and connected sequentially, sharing the same fronthaul link through a cable or stripe, effectively serving as a bus for data transfer, power supply, and synchronization among the APs. They differ from CF in three main aspects. RSs avoid the need for dedicated fronthaul links between each AP and the corresponding processing unit. RSs are not tied to the coverage tier of a network; instead, they are dedicated to increasing the capacity in specific small-to-moderate areas where deploying conventional BSs is challenging. RSs are built with handset-like hardware, meaning that each stripe can have an even larger number of antennas, which can operate phase-coherently. This technology is being adopted by Ericsson and may be considered as a single-dimension large intelligent surface (LIS) \cite{LIS_Hu}, where the antennas are serially embedded within flexible stripes that contain all the radio frequency (RF) components, including power amplifiers, phase shifters, filters, modulators, analog-to-digital converter and digital-to-analog converter. Despite their promising potential, several research challenges are present. A subset of them involves the development of the algorithms needed for resource allocation (RA) in a distributed system, which can be inherently complex, necessitating the development of efficient algorithms capable of real-time operation. This is also justified by the constraints of backhaul and fronthaul links to ensure low latency and real-time communication between the APs and the CPU.

\subsection{Related Work}

The RS network was proposed in \cite{RS_Interdonato}, in which the authors managed to show the advantages of CF mMIMO while addressing the practical deployment issues of this architecture regarding the fronthaul capacity. Following this paper, the literature has been focused on the development of sequential processing algorithms for the uplink (UL) \cite{RS_Shaik, RS_Ma, RS_Shaik2, RS_Yuan, RS_Martins} and downlink (DL) \cite{RS_Miretti} in an RS scenario. The authors of \cite{RS_Shaik2} developed a UL sequential linear processing (SLP) algorithm, known as optimum sequence linear processing (OSLP), which was proved to be optimal in both the maximization of the SE and the minimization of the mean square error (MSE). This approach is capable of achieving the same performance as the optimal centralized implementation based on linear minimum MSE (LMMSE) combining proposed in \cite{MMSE_Bjornson} for all SINRs. However, it does so in a distributed manner, thereby significantly reducing the fronthaul requirements usually observed in CF mMIMO networks. 

However, different use cases are also being studied with RSs. In \cite{RS_Chiotis}, the authors investigated the maximization of the per-UE UL SE in an RS with limited capacity with a novel arrangement of APs. They used a compare-and-forward strategy to enable clustering cooperation. In \cite{RS_Conceicao}, the authors proposed different approaches to optimize the UL power distribution in an RS network. Their objective was to ensure complete fairness among UEs by maximizing the minimum per-UE SE. This work was extended to a bi-objective case in \cite{Bi_Conceicao}, where the optimization of the UL power vector considers not only the max-min fairness but also the max-sum SE. 

Another important CF mMIMO solution that network engineers have been often considering is the UE-centric operation \cite{CF_Buzzi}. In this scenario, only a subset of APs are in charge of serving a specific UE. This not only alleviates the signal processing load and computational complexity of the AP processing unit but also has the potential to increase the SINR, as only a fraction of UEs are now included in the interference term. Multiple works have already been investigating an AP selection scheme for CF. In \cite{APS_Dao}, the authors proposed an AP selection scheme by leveraging large-scale fading and calculating the effective channel gain from all UEs to all APs, along with considering the channel quality of each UE. In \cite{APS_Biswas}, an AP selection is proposed based on a clustering approach aimed at reducing computational complexity and mitigating pilot contamination which is achieved through the utilization of a machine learning (ML) algorithm. In \cite{APS_Wang}, an AP selection approach is investigated under a Rician fading channel. In \cite{APS_Wei}, the authors studied the AP selection problem using game theory and modeled the formation a UE-centric AP service cluster as a local altruistic game. In \cite{APS_Jung}, the authors prioritize the DL SE of the UEs with poor channel conditions by employing an AP selection scheme using a linear assignment algorithm. In \cite{APS_Ranasinghe}, the authors implemented a graph neural network-based AP selection scheme for CF mMIMO. This approach is capable of accurately predicting the links between the UEs and APs, outperforming proximity-based AP selection schemes. Furthermore, it does not require the knowledge of the received signal strengths of all the neighbouring APs, unlike the large-scale fading coefficient-based AP selection schemes. A lower complexity approach compared to the MMSE-based AP selection is presented in \cite{APS_Shakya}, where the CPU uses channel estimation (CE) to perform an AP selection scheme by measuring the SINR for each UE. 

In \cite{APS_Mendoza}, a deep reinforcement learning-based (DRL) approach is employed to select the APs serving each UE given its position. The aim of this approach is to study the trade-off between quality of service (QoS) and power consumption. In \cite{APS_Vu}, the authors developed a joint power allocation and AP selection strategy to minimize the total energy consumption of a CF mMIMO system with QoS constraints. In \cite{APS_OHurley}, the authors employed an AP selection scheme with a full-pilot ZF combining approach, which offered reasonably good SE while increasing the system's PE. In \cite{APS_Jiang}, a distanced-based AP selection approach is combined with an orthogonal assignment of subcarriers among different UEs to enhance the power and SEs of a CF mMIMO-OFDM system. \cite{APS_Ghiasi} leveraged a DRL approach to improve the PE of a CF mMIMO system by determining the optimal AP-UE association subject to minimum rate constraints for all devices. A different approach is proposed in \cite{APS_D'Andrea} that groups the APs in cell-centric clusters connected to different cooperative CPUs. However, the AP selection scheme proposed in the paper allowed the association of each UE to a virtual cluster. This AP-UE association method aimed at the maximization of the total sum-rate and offered a considerably lower backhaul load when compared to full CF approaches. The authors in \cite{APS_Wang2} combine beamforming and an AP selection scheme for the DL of a mmWave CF mMIMO system. In \cite{APS_Wang3}, the authors aimed to design AP-UE association, hybrid beamforming, and fronthaul compression in a mmWave CF mMIMO system. In \cite{APS_Conceicao}, the authors developed a centralized AP selection scheme for the UL of an RS network. This scheme incorporates two sequential equalization methods: OSLP and maximum ratio combining (MRC). Initially, the aim was to solve a single-objective optimization problem to enhance the UL sum rate for UEs in terms of SE. Then, the focus was on improving a bi-objective operation of the overall RS network, by enhancing the SE and ensuring load balancing among APs.

\subsection{Contributions}

Most of the papers mentioned in the previous section focused on proposing AP selection schemes (possibly combined with other optimization problems) to improve SE, PE, computational complexity, and/or pilot contamination in the UL or DL. However, since most AP selection algorithms are developed to be compatible with CF networks, these are implemented in a centralized strategy in which a CPU is responsible for maintaining synchronization and calculating the optimized AP-UE association solution. Due to the high concentration of fronthaul links and implementation costs, alternative AP selection schemes should be adapted to UE-centric models such as the RS architecture concept and, thus, implemented through a distributed or sequential implementation strategy. Consequently, since we aim to move a significant part of the signal processing load from the CPU to the APs, a low-complexity and efficient optimization method should be envisioned.

In this work, we aim to implement an RS network with a UE-centric approach by developing sequential and parallel AP selection schemes. Additionally, these schemes determine, for each UE, which antenna or antennas of a single AP should take part in the reception of its UL signal. The first AP selection strategy is performed sequentially for each AP within the AP's local processing unit (LPU) in order to maximize the total UL SE regarding all UEs. The second strategy enables the simultaneous allocation of APs, operating in parallel, without necessitating any exchange of information. Given the efficiency of the approach developed in \cite{APS_Conceicao}, we adopted this centralized MRC-based detection along with the GA RA algorithm whose aim was to improve the total sum SE as a benchmark. Since centralized OSLP (COSLP) requires constant knowledge of the UL equalization coefficients used across all APs, any AP selection scheme incorporating the OSLP can only be executed in the CPU, i.e. it must be centralized. However, each centralized AP selection algorithm has high computational complexity and significant fronthaul signalling requirements, which affect network scalability due to the large solution search space and centralized execution. Therefore, we aim to implement this AP decentralized selection scheme in both sequential and parallel models to increase speed convergence and reduce complexity and capacity requirements. To achieve these objectives, the AP selection schemes developed here are designed to be both efficient and fast.

Furthermore, we aim to evaluate the adaptability of the developed AP selection schemes when the network undergoes configuration changes, such as variations in the number of APs and UEs. To do this, we initialize the optimization solving algorithm with the previous AP-UE association matrix (the one without any UE added or removed) and compare the final solution with a case where there is no previously known solution. Since the main motivation for the development of these AP selection strategies is to reduce the computational complexity, and given that our potential solutions are binary matrices, we leverage a meta-heuristic (MH) model known as genetic algorithm (GA) \cite{GA_Banzhaf, APS_Conceicao}, which works with a population of solutions. To the best of the authors’ knowledge, this is the first work aiming to adapt an AP selection scheme to the sequential nature of the RS network. Thus, the two main contributions of this paper are as follows:

\begin{itemize}

    \item We propose two AP selection schemes with sequential and parallel strategies, respectively. Considering an RS network with $K$ UEs and $L$ APs, each equipped with $N$ antennas, in both cases, each AP's LPU independently performs the selection/allocation procedure locally to determine the subset of UEs to serve (from the $K$ UEs) and antenna(s) allocated to each of them. In the sequential approach, the maximization of SE at AP $l$ considers all APs $1$ through $l{-}1$. SE exhibits a monotonous convergence towards the SE obtained with centralized approaches. In the parallel scheme, optimization occurs concurrently among APs, each sharing only its local view with its neighbors. This happens whenever changes in the network are sensed, such as new inputs from neighboring APs or changes in the number of UEs. Both approaches involve considering a sub-matrix of dimensions $K \times N$.

    \item We evaluate the adaptability of the centralized, sequential, and parallel MRC (CMRC, SMRC and PMRC) strategy optimization algorithm for two different network instances: when a new UE is added in a random position in the RS network (having now $K{+}1$ UEs) and when a random UE is removed from it (resulting in $K{-}1$ UEs). To measure the performance of the optimization algorithm and compare it with the original scenario where we have $K$ UEs, we consider two cases: one where the initial solutions are randomly selected and one where we include the original AP-UE association matrix.

\end{itemize}

To demonstrate the validity and applicability of our results, the additional secondary contributions are:

\begin{itemize}

    \item The development of a low-complexity optimization MH algorithm based on a population-based GA model. We present a detailed description of its computational complexity and pseudo-code.

    \item The comparison of the performance of the proposed approaches with a centralized strategy of the AP selection algorithm proposed in \cite{APS_Conceicao}. This is used as the benchmark solution, in which the CPU computes an efficient solution for the association between all APs and UEs in the network. This strategy considers the full $K \times LN$ AP-UE association matrix in the optimization method for both MRC and OSLP.

    \item The evaluation of the robustness of the centralized, sequential, and parallel strategy optimization approaches by considering several network scenarios with different AP to UE densities.
    
\end{itemize}

\subsection{Outline and Notation}

The outline of this work is as follows. Section II reviews the main concepts of the RS network including the pilot transmission, CE and UL detection. Section III formalizes the centralized, sequential, and parallel AP selection optimization which aims to maximize the total UL SE. This section also presents the GA which was used as an MH tool to solve the AP selection problem along with its computation complexity and the fronthaul signalling of both schemes. Section IV includes the simulation results comparing the AP selection schemes with both the MRC and OSLP UL equalization. Finally, section V presents the main conclusions.

Throughout this paper the following notation will be employed: lower-case bold lettering (e.g., $\textbf{s}$) denotes a block/vector of samples at the time domain, while non-bold lower-case lettering (e.g., $s$) is used to denote the symbols/samples of each of those block/vectors. Furthermore, the superscripts $\textbf{A}^T$ and $\textbf{A}^H$ denote transpose and hermitian of the matrix $\textbf{A}$, respectively. Additionally, $\text{tr}\left(\textbf{A}\right)$ denotes the trace of $\textbf{A}$ and $\text{diag}\left(\textbf{a}_1,\cdots,\textbf{a}_n\right)$ is a $n$-squared diagonal matrix whose non-zero components are $\textbf{a}_1,\cdots,\textbf{a}_n$. $\textbf{I}_n$ is a $n \times n$ identity matrix, $\mathbb{E}\left\{\right\}$ is the expected value operation and $\mathcal{N}_\mathbb{C}(0,\textbf{R})$ is a circularly symmetric complex Gaussian distribution with correlation matrix $\textbf{R}$. Additionally, $\textbf{A}_1 \odot \textbf{A}_2$ is the Hadamard product between matrices $\textbf{A}_1$ and $\textbf{A}_2$, resulting in a same-sized matrix with multiplied corresponding elements. Table \ref{table_parameters} presents all important parameters included in this paper.

\begin{table*}[t]%
\centering
\caption{All parameters and their meaning}
\label{table_parameters}
\resizebox{2\columnwidth}{!}{\begin{tabular}[t]{|c l | c l |}
\hline
  \textbf{Parameter} & \textbf{Meaning} & \textbf{Parameter} & \textbf{Meaning}\\
\hline
  $L$ & Total number of APs & $\hat{\boldsymbol{\mathcal{H}}}_{kl}$ & Augmented CE matrix \\
  $K$ & Total number of UEs & $\textbf{K}_l$ & Correlation matrix of the augmented error estimation plus AWGN \\
  $\tau_c$ & Channel's coherence block duration & $N_m$ & Maximum number of iterations \\
  $\tau_p$ & Total number of UL orthogonal pilots & $N_i$ & Maximum number of iterations without enhancement in the objective function  \\
  $\tau_u$ & Total number of UL payload symbols & $N_{pop}$ & Size of the population \\
  $N$ & Total number of antennas in an AP & $n_{iter}$ & Number of required iterations\\
  $\textbf{h}_{kl}$ & Channel between UE $k$ and AP $l$ & $d_{kl}$ & Distance between UE $k$ and AP $l$\\
  $\textbf{R}_{kl}$ & Correlation matrix between the AP's antennas for UE $k$ & $f_c$ & Carrier frequency \\
  $\beta_{kl}$ & Large-scale fading coefficient between UE $k$ and AP $l$ & $h_{AP}$ & AP's antenna height\\
  $d_H$ & Spacing between the AP's antennas & $h_{UE}$ & UE's antenna height\\
  $\lambda$ & Wavelength & $\nu_{ki}$ & Distance between UE $k$ and UE $i$ \\
  $F_{kl}$ & Shadow fading factor from UE $k$ and AP $l$ & $\varrho_{lj}$ & Distance between AP $l$ and AP $j$ \\
  $\sigma_{sh}^2$ & Variance of the shadow fading factor & $\Delta$ & AP/UE shadowing contributions \\
  $B$ & Channel's bandwidth & $d_c$ & Decorrelation distance from shadow fading model \\
  $\sigma^2$ & AWGN power & $D$ & Side of a squared area \\
  $\hat{\textbf{h}}_{kl}$ & MMSE CE of $\textbf{h}_{kl}$ & $\sigma_f$ & Noise figure \\
  $\textbf{y}_l$ & UL signal received at AP $l$ & $k_B$ & Boltzmann constant \\
  $p_k$ & UL power transmitted by UE $k$ & $T_0$ & Noise temperature \\
  $P_{max}$ & Maximum transmission power for UEs & $d_{sh}$ & Distance included in the shadow fading model \\
  $\textbf{n}_l$ & AWGN at the receiver of AP $l$ & $N_a$ & Independent attempts of GA \\
  $\tilde{\textbf{h}}_{kl}$ & CE error of $\textbf{h}_{kl}$ & $\textbf{D}^{(t,i)}$ & $i$th solution of $\textbf{D}$ at generation $t$ \\
  $\textbf{H}_l$ & UL channel matrix from AP $l$ to all UEs & $\hat{\textbf{D}}^{(t,i)}$ & $i$th solution in the crossover population at generation $t$ \\
  $\textbf{s}$ & Combined UL signal at AP $l$ & $\tilde{\textbf{D}}^{(t,i)}$ & $i$th solution in the offspring population at generation $t$ \\
  $\hat{\textbf{H}}_l$ & Matrix of MMSE CE for the UL & $\textbf{D}^{(t,i)}_l$ & Sub-matrix of $\textbf{D}^{(t,i)}$ for AP $l$ \\
  $\tilde{\textbf{H}}_l$ & Matrix of CE error for the UL & $\boldsymbol{\mathcal{D}}^t_l$ & Sub-population of $\boldsymbol{\mathcal{D}}^t$ for AP $l$ \\
  $\textbf{w}_l$ & Vector with CE errors and the AWGN in the UL & $\tilde{\boldsymbol{\mathcal{D}}}^t$ & Offspring population at generation $t$ \\
  $\boldsymbol{\Sigma}_l$ & Correlation matrix of $\textbf{w}_l$ & $t_k$ & Size of the tournament in GA \\
  $\textbf{D}$ & AP selection matrix & $\tilde{\boldsymbol{\mathcal{D}}}^t_l$ & Sub-population of $\tilde{\boldsymbol{\mathcal{D}}}^t$ for AP $l$ \\
  $\textbf{D}_l$ & Sub-matrix of $\textbf{D}$ for AP $l$ & $\textbf{b}^i$ & Binary mask for individual $i$ \\
  $\textbf{D}_{kl}$ & Sub-matrix of $\textbf{D}_l$ for UE $k$ & $\textbf{b}^i_l$ & Sub-matrix of $\textbf{b}^i$ for AP $l$ \\
  $\textbf{E}_{kl}$ & Matrix related to the AP selection scheme & $p_r$ & Crossover probability in GA \\
  $\hat{\textbf{s}}_l$ & Estimation of $\textbf{s}_l$ & $\hat{\boldsymbol{\mathcal{D}}}^t$ & Population resulting from crossover at generation $t$ \\
  $\textbf{A}_l$ & Matrix of AP $l$ allowing sequential estimation & $\hat{\boldsymbol{\mathcal{D}}}^t_l$ & Sub-population of $\hat{\boldsymbol{\mathcal{D}}}^t$ for AP $l$ \\
  $\textbf{B}_l$ & Matrix of AP $l$ allowing local estimation & $p_m$ & Mutation probability in GA \\
  $\textbf{z}_l$ & Augmented received signal at AP $l$ & $p_{rm}$ & Increase rate of $p_r$ \\
  $\bar{\textbf{B}}_l$ & Augmented receiver combining matrix of AP $l$ & $p_{mm}$ & Increase rate of $p_m$ \\
  $\textbf{T}_l$ & Matrix employed in OSLP detection & $p_{ri}$ & Nominal value of $p_r$ \\
  $\textbf{P}_l$ & Matrix used in OSLP detection & $p_{mi}$ & Nominal value of $p_m$\\
  $\text{SE}_{kl}$ & SE of UE $k$ at AP $l$ & $p_{rmax}$ & Maximum value of $p_r$ \\
  $\text{SINR}_{kl}$ & SINR of UE $k$ at AP $l$ & $p_{mmax}$ & Maximum value of $p_m$\\
  $\boldsymbol{\mathcal{B}}_{lk}$ & Sub-matrix of $\bar{\textbf{B}}_l$ & $F_a^b$ & Total fronthaul signaling for detection scheme $a$ implemented in $b$ manner \\
  $D_{fd}$ & Fraunhofer distance & $\sigma_{\phi}$ & Std of the angular deviations\\
\hline
\end{tabular}}
\end{table*}

\section{Radio Stripe Network}

We consider an RS network implemented in a single stripe that contains $L$ sequentially located APs, each one equipped with a linear array of $N$ antennas. This set of APs is in charge of serving $K$ single-antenna UEs that are located within a square area in the same time-frequency resources.

\subsection{Pilot Transmission and Channel Estimation}

The CE phase is performed through the transmission of pilot sequences in the UL. We adopt a time-frequency block of $\tau_c$ samples in which the channel between AP $l {=} 1,\cdots, L$ and UE $k {=} 1,\cdots, K$, $\textbf{h}_{kl}$ is constant \cite{MIMO_Bjornson, MMSE_Bjornson}. Each channel follows a correlated Rayleigh fading distribution $\textbf{h}_{kl} \in \mathbb{C}^N \sim \mathcal{N}_\mathbb{C}(\textbf{0},\textbf{R}_{kl})$, where $\textbf{R}_{kl} \in \mathbb{C}^{N \times N}$ is the spatial correlation matrix between AP $l$ and UE $k$ and the large fading coefficient is $\beta_{kl} {=} \frac{\text{tr}(\textbf{R}_{kl})}{N}$.

Because of this we consider a UL time-division-duplex (TDD) frame of $\tau_c$ samples divided in $\tau_c = \tau_p + \tau_u$. The first term is used for CE, $\hat{\textbf{h}}_{kl}$, in which one pilot signal from the orthogonal basis is assigned to UE $k$ and transmitted to all APs with power $P_{max}$. The MMSE CE approach from \cite{MIMO_Bjornson} is employed in this paper. The second term is used for UL payload transmission and usually $\tau_p \le K$ and, consequently, $\tau_u \gg \tau_p$. Since the number of UEs in the system typically exceeds the available mutually orthogonal pilot sequences transmitted in the UL, pilot contamination interference arises in the CE vectors. The extent of this interference is influenced by the number of samples allocated for CE, denoted as $\tau_p$.

\subsection{Uplink Signal Reception}

\begin{figure}[t!]
\centering
\includegraphics[width=1\columnwidth]{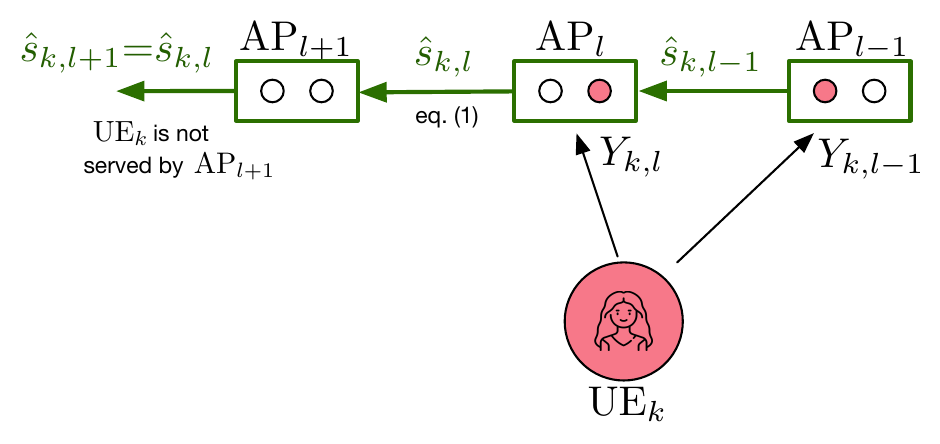}
\caption{Depiction of a general UL SLP algorithm in an RS network.}
\label{fig:RS_SLP}
\end{figure}

In the UL channel, the UE signals are transmitted to all APs which cooperate sequentially through a linear processing algorithm to estimate the original signal. This process is illustrated in Fig. \ref{fig:RS_SLP}.  The AP selection scheme denoted by the red color in the APs antennas can be included in the equalization estimation by defining a Boolean matrix $\textbf{D} \in \mathbb{B}^{K \times LN}$. We consider that $\textbf{D}{=}\left[\textbf{D}_1, \cdots, \textbf{D}_l, \cdots, \textbf{D}_L\right]$, where $\textbf{D}_l \in \mathbb{B}^{K \times N}$ and $\textbf{D}_l{=}\left[\textbf{D}_{1l},\cdots,\textbf{D}_{kl}, \cdots, \textbf{D}_{Kl}\right]$, where $ \textbf{D}_{kl} \in \mathbb{B}^{N}$.

From Fig. \ref{fig:RS_SLP}, AP $l$ is able to estimate the $K$-sized signal estimated vector, $\hat{\textbf{s}}_l {=} \left[\hat{s}_{1l},\cdots,\hat{s}_{Kl}\right]^T$, by receiving not only the estimation from AP $l{-}1$, $\hat{\textbf{s}}_{l{-}1}$ but also the local received signal given by
\begin{equation}  \label{eq:ReceivedSignal}
    \textbf{y}_l = \left(\textbf{D}_l^T \odot \hat{\textbf{H}}_l\right)\textbf{s} + \textbf{w}_l,
\end{equation}
where $\textbf{H}_l {=} \left[\textbf{h}_{1l},\cdots,\textbf{h}_{Kl} \right] \in \mathbb{C}^{N \times K}$ is the channel matrix for AP $l$ and consequently $\hat{\textbf{H}}_l$ is its CE matrix, $\textbf{s} {=} \left[s_1,\cdots,s_K \right]^T \in \mathbb{C}^{K}$ is the UL signal vector with power $p_k{=}P_{max}$ and $\textbf{w}_l {=} \left(\textbf{D}_l^T \odot \tilde{\textbf{H}}_l\right)\textbf{s} {+} \textbf{n}_l$ includes $\tilde{\textbf{H}}_l {=} \textbf{H}_l {-} \hat{\textbf{H}}_l$ and the additive white Gaussian noise (AWGN) $\textbf{n}_l \sim \mathcal{N}{_\mathbb{C}}\left(0,\sigma^2\textbf{I}_N\right)$. The latter term follows $\textbf{w}_l \sim  \mathcal{N}{_\mathbb{C}}\left(0,\boldsymbol{\Sigma}_l\right)$ where
\begin{equation}  \label{eq:CovarianceNoiseError}
    \boldsymbol{\Sigma}_l = \sum_{i=1}^K p_i \left(\textbf{E}_{il} \odot \tilde{\textbf{R}}_{il} \right) + \sigma^2\textbf{I}_N.
\end{equation}

The matrices $\textbf{E}_{kl} \in \mathbb{B}^{N \times N}$ adjust the channel spatial correlation of $\tilde{\textbf{h}}_{kl} {=} \textbf{h}_{kl} {-} \hat{\textbf{h}}_{kl}$ according to the AP selection scheme given by $\textbf{D}$. The $\hat{\textbf{s}}_l$ vector from Fig. \ref{fig:RS_SLP} can be estimated by \cite{RS_Shaik,     RS_Shaik2}
\begin{equation} \label{eq:SignalGeneralized}
    \hat{\textbf{s}}_l = \textbf{A}_l\hat{\textbf{s}}_{l{-}1} + \textbf{B}_l\textbf{y}_l = \bar{\textbf{B}}_l\textbf{z}_l,
\end{equation}
where two weighting matrices $\textbf{A}_l \in \mathbb{C}^{K \times K}$ and $\textbf{B}_l \in \mathbb{C}^{K \times N}$ are included. The term $\textbf{z}_l {=} \left[\textbf{y}_1^H,\cdots,\textbf{y}_l^H\right]^H$ includes all the received signal vectors from AP $1$ to $l$ and
\begin{equation} \label{eq:AugmentedMatrix}
    \bar{\textbf{B}}_l = \left\{
   \begin{array}{lcl}
   
      \left[\textbf{A}_l\bar{\textbf{B}}_{l-1} \quad \textbf{B}_l\right], & {\text{if }l > 1}, \\
      \textbf{B}_1, & {\text{if } l = 1}.
    \end{array}
    \right. \,
\end{equation}   

The MRC scheme is modeled by 
\begin{equation} \label{eq:MRC_combining}
    \left\{\begin{array}{lcl}
        \textbf{A}_l {=} \textbf{I}_K, \\
        \textbf{B}_l {=} \hat{\textbf{H}}_l^H,
    \end{array}
   \right. \,
\end{equation}
and the OSLP scheme \cite{RS_Shaik2} is given by 
\begin{equation} \label{eq:OSLP_combining}
    \left\{\begin{array}{lcl}
    \textbf{A}_l {=} \textbf{I}_K  {-} \textbf{T}_l\hat{\textbf{H}}_l, \\
    \textbf{B}_l {=} \textbf{T}_l, 
    \end{array}
    \right.
\end{equation}
where 
\begin{equation} \label{eq:OSLP_combining2}
    \textbf{T}_l {=} \textbf{P}_{l{-}1}\hat{\textbf{H}}_l^H\left(\boldsymbol{\Sigma}_l {+} \hat{\textbf{H}}_l\textbf{P}_{l{-}1}\hat{\textbf{H}}_l^H\right)^{-1},
\end{equation}
and
\begin{equation} \label{eq:OSLP_combining3}
    \textbf{P}_{l{-}1} {=} \left(\textbf{I}_K {-} \textbf{T}_{l{-}1}\hat{\textbf{H}}_{l{-}1}\right)\textbf{P}_{l{-}2}.
\end{equation}

The SE of UE $k$ in AP $l$ is calculated by 
\begin{equation} \label{eq:SE}
    \text{SE}_{kl} = \left(1 - \frac{\tau_p}{\tau_c}\right)\mathbb{E}\left\{\log_2\left(1 + \text{SINR}_{kl}\right)\right\},
\end{equation}
where
\begin{equation} \label{eq:SINR}
    \text{SINR}_{kl} = \frac{p_k\left|\boldsymbol{\mathcal{B}}_{lk}^H \hat{\boldsymbol{\mathcal{H}}}_{kl}\right|^2}{\sum_{i=1, i \neq k}^{K} p_i\left|\boldsymbol{\mathcal{B}}_{lk} \hat{\boldsymbol{\mathcal{H}}}_{il}\right|^2 + \boldsymbol{\mathcal{B}}_{lk}^H\textbf{K}_l\boldsymbol{\mathcal{B}}_{lk}},
\end{equation}
and $\hat{\boldsymbol{\mathcal{H}}}_{kl}{=} \left[\textbf{D}_{k1} \odot \hat{\textbf{h}}_{k1},\cdots,\textbf{D}_{kl} \odot \hat{\textbf{h}}_{kl}\right]^T \in \mathbb{C}^{Nl \times 1}$, $\boldsymbol{\mathcal{B}}_{lk} \in \mathbb{C}^{Nl \times 1}$ such that $\bar{\textbf{B}}_l = \left[\boldsymbol{\mathcal{B}}_{l1}, \cdots, \boldsymbol{\mathcal{B}}_{lK}\right]^T$ and $\textbf{K}_l = \text{diag}\left(\boldsymbol{\Sigma}_1,\cdots,\boldsymbol{\Sigma}_l\right)$.

\section{AP Selection Optimization}

In this section, we will provide a detailed description of the centralized, sequential, and parallel AP selection strategies.

\subsection{Centralized Implementation}

In this approach, we consider that the UL detection is adapted to the RS architecture by being sequential (Fig. \ref{fig:RS_SLP}) but the AP selection is formalized through an optimization problem which is entirely solved in the CPU, as in \cite{APS_Conceicao}. This means that the AP-UE association matrix, $\textbf{D}$, is fully considered in the optimization procedure. Since we must know the whole AP-UE association matrix of all APs to solve the problem, the optimization algorithm must be implemented in the CPU at a stage that is earlier than the UL signal reception phase. However, its solutions are valid through an entire coherent time-frequency block of $\tau_c$ samples. The solutions are the different versions of $\textbf{D}$ and the total search space includes $2^{KLN}$ possible solutions. The objective function of this problem is the total SE of the network, which is measured at the last AP or at the CPU, i.e. $l{=}L$. The problem can be formulated by $f_1\left(\textbf{D}\right)$

\begin{equation} \label{eq:MaxSum}
\begin{aligned}
\mathbb{P}_1: \max_{\textbf{D}} \quad & f_1\left(\textbf{D}\right) =\sum_{k=1}^{K}\text{SE}_{kL}. \\
\end{aligned}
\end{equation}

Since this strategy is aimed at maximizing the UL performance of the RS network, we can consider both the CMRC and COSLP detection schemes \cite{RS_Shaik} to calculate the total throughput in Eq. \eqref{eq:MaxSum}. By looking at Eqs. \eqref{eq:MRC_combining} and \eqref{eq:OSLP_combining}, it can be argued that both approaches are viable in a centralized AP selection scheme. As a matter of fact, for the MRC, each AP performs the UL detection by leveraging only the local CE vector. Thus, when we are employing the AP selection optimization problem with this detection scheme and the $n$th antenna of AP $l$ is set to $1$ for UE $k$, there is no need for sharing extra information with any other AP. In other words, when that given index is set to $1$, the AP $l$ is responsible for serving UE $k$ through the $n$th antenna but since MRC only requires local knowledge of the CE, this does not affect the matrices defined in Eq. \eqref{eq:MRC_combining} for other APs. All of this can be easily managed in the CPU. On the other hand, for the OSLP, we can observe through Eqs. \eqref{eq:OSLP_combining2} and \eqref{eq:OSLP_combining3} that the way each AP performs the UL detection depends not only on local CE but also on previous APs' CE information. This contradicts what had been stated for MRC: when the $n$th antenna of AP $l$ is set to $1$ for UE $k$, this affects not only the local AP UL detection but also the subsequent detection of APs $l{+}1,l{+}2,\cdots,L$. In other words, when this happens, AP $l$ is in charge of serving UE $k$ through the $n$th antenna and there must be an exchange of information between AP $l$ and the following APs because this affects their receiver OSLP coefficients, changing the matrices defined in Eq. \eqref{eq:OSLP_combining}. However, we are solving the AP selection optimization problem in the CPU, which has full access to this information at any time.

\subsection{Sequential Implementation}

On the other hand, in this approach, both the UL detection and the AP selection optimization procedures are performed sequentially. This means that the AP-UE association matrix is iteratively optimized in each AP. In other words, in AP $l$, an optimization algorithm is performed in its LPU to find the best RA instance for the UEs through sub-matrix $\textbf{D}_l$. The best possible solution in the first loop of RA is the one that maximizes the UL SE considering the contributions of APs $1,\cdots,l$. Once all the APs have an AP-UE association matrix, we can consider their SE contributions in AP $l$, despite only optimizing their local sub-matrix. Since each AP performs a local optimization, it only requires its local CE vector along with its AP-UE association sub-matrix, $\textbf{D}_l$. The optimized solution is also valid through a coherent time-frequency block. Although requiring at least $L$ iterations in order to compute the whole $\textbf{D}$ matrix, the local optimization approach has a much smaller solution search space totalizing $2^{KN}$ solutions for each sub-matrix $\textbf{D}_l$. The objective function of this iterative problem depends on whether the AP selection is being performed on the first loop of the RA or not. In the first case, it is defined by the total SE given by APs $1$ to $l$ while in the second one, it is defined by the total SE given by all $L$ APs. The problem can be formulated by $f_2\left(\textbf{D}_l\right)$

\begin{equation} \label{eq:MaxSum_local}
\resizebox{1\columnwidth}{!}{$
\begin{aligned}
\mathbb{P}_2: \max_{\textbf{D}_l} \quad & f_2\left(\textbf{D}_l\right) = \left\{\begin{array}{lcl} 
\sum_{l'=1}^{l} \sum_{k=1}^{K}\text{SE}_{kl'}, \text{in the first loop of RA}, \\
\sum_{i=1}^{L} \sum_{k=1}^{K}\text{SE}_{kl}, \text{otherwise}.
\end{array}
\right. \,
\\
\end{aligned}$}
\end{equation}

Unlike the centralized approach, where we consider that the CPU has access to all UL information, in the sequential approach each AP can only perform the AP selection optimization with its local information. Although it is possible for the APs to share their local information (such as the CE vectors), through Fig. \ref{fig:RS_SLP}, in this approach we aim at a decentralized and CPU-independent approach, in which the AP selection is performed in their LPUs. Following the explanation from the previous subsection, we can state that the only combining scheme that is viable in a sequential AP selection approach is the MRC owing to the fact that by looking at Eq. \eqref{eq:MRC_combining}, the equalization matrices for AP $l$ are only defined by its local CE. Furthermore, the employment of the OSLP detection scheme with the sequential AP selection approach is unfeasible since that would require the LPUs of APs $l$ and $l{+}1,l{+}2,\cdots,L$ to share their local CEs in order to define the equalization matrices in Eq. \eqref{eq:OSLP_combining}.

\subsection{Parallel Implementation}

Similarly to the sequential implementation, the parallel approach aims at the optimization of the AP-UE association matrix, which is done in each AP, through sub-matrix $\textbf{D}_l$. However, this process is done simultaneously, i.e. in parallel, and, therefore, each AP $l$ maximizes the UL SE discarding any contributions from previous APs $1,\cdots,l{-}1$. This causes the SE values to be wrongfully estimated by the AP selection optimization scheme, as all of the inter-UE interference resulting from the allocation of the UEs to different APs is ignored. Nevertheless, the main purpose of the parallel implementation is to test whether the AP-UE association matrix output by this approach is still able to provide efficient SE results while significantly reducing the time complexity of the sequential algorithm. The problem can be formulated by $f_3\left(\textbf{D}_l\right)$

\begin{equation} \label{eq:MaxSum_parallel}
\begin{aligned}
\mathbb{P}_3: \max_{\textbf{D}_l} \quad & f_3\left(\textbf{D}_l\right) = \sum_{k=1}^{K}\text{SE}_{kl}, l = 1,\cdots,L.
\end{aligned}
\end{equation}

Once again, the parallel approach relies on local AP information. Thus, we can only include the MRC scheme.

\subsection{Genetic Algorithm}

The primary rationale for employing an MH algorithm lies in its ability to substantially simplify the process of obtaining (near-)optimal solutions within a time frame suitable for the RS system's operation. While deep learning-based approaches offer low time-complexity during execution, they require an extensive offline training phase, which can be time-consuming and necessitates the creation of a large training dataset tailored to a specific network scenario. In contrast, MH algorithms can be executed in the network’s scheduling period, providing adaptive, much closer to real-time solutions without the need for prior training or scenario-specific adaptation.

As a result, MH algorithms emerge as viable candidates for delivering (near-)optimal performance with reasonable computation times and complexity, making them practical, scalable, and efficient optimization techniques for deployment across various RS networks. We leverage the GA MH algorithm as the optimization variables in this optimization are binary.

The GA \cite{GA_Banzhaf, GA_Michalewicz} MH algorithm involves a set of potential solutions that are transformed over successive iterations and whose aim is to find a high-quality solution, ideally the optimal solution, for a specific optimization problem that is being solved. The set of potential solutions is denoted by population, each potential solution is known as an individual or chromosome and a certain value of a given decision variable in an individual is known as a gene. Additionally, the iterations of the optimization process are denoted as generations and the sequence of the various populations reflects the evolution over time. In this work, the objective function is given by Eqs. \eqref{eq:MaxSum}, \eqref{eq:MaxSum_local}, or \eqref{eq:MaxSum_parallel} depending if we aim to solve the centralized, sequential, or parallel AP selection optimization problem. Furthermore, the population is defined by a set of $N_{pop}$-sized AP-UE association matrices given by $\textbf{D}$. The transformations that are applied to each solution in a given generation are known as crossover and mutation. The crossover trades some of the genes between two previously selected individuals known as parents in order to create two new child individuals while the mutation adds diversity to the population by randomly changing a single gene from an individual. The primary operations of the proposed algorithm are based on the GA framework. However, to add a higher degree of diversity in the populations, the offspring population is generated to be three times the size of the original population. This was achieved by performing a double crossover between the two parent populations, resulting in two distinct offspring populations. Additionally, we incorporated the current population into the offspring population.

\subsubsection{Centralized vs Sequential vs Parallel Approaches} In the centralized and parallel approaches, we firstly assume that the population is randomly initialized, being defined by $\boldsymbol{\mathcal{D}}^t = \left\{\textbf{D}^{(t,i)}, i = 1,\cdots,N_{pop}\right\}$, where $t$ is the index of the generation and $i$ is the index of each individual. The stopping requirements for the centralized scheme are a maximum number of generations, $N_m$, or a maximum number of iterations, $N_i$, without any improvement on the objective function within a certain tolerance, $\varepsilon$. However, in the sequential approach, we can only randomly initialize the first sub-matrix of the $N_{pop}$ $\textbf{D}^{1,i}$ individuals in $\boldsymbol{\mathcal{D}}^1$, i.e. $\textbf{D}_1^{1,i}$. Owing to this, the stopping criteria for this approach are a maximum number of generations, $LN_m$, or a maximum number of iterations, $LN_i$, without any improvement on the objective function with the same tolerance. In the parallel approach, the stopping requirements are a maximum number of generations, $N_m$, or a maximum number of iterations, $N_i$, in which all APs output a solution without any improvements on the objective function with a similar tolerance. The following GA operations are the same for all approaches. Nevertheless, we should mention that in the centralized approach all genes from the $\textbf{D}^{(t,i)}$ individuals are accessible in generation $t$ and can suffer modifications from the genetic operators. In the parallel scheme, in generation $t$, each LPU from AP $l$ performs the GA operations based only on sub-matrix $\textbf{D}_l^{(t,i)}$ from $\textbf{D}^{(t,i)}$. In the sequential approach, only the sub-matrix of AP $l$, $\textbf{D}_l^{(t,i)}$, of the $\textbf{D}^{(t,i)}$ individuals is accessible in generation $t$. Because of this we defined the sub-population of $\boldsymbol{\mathcal{D}}^t_l = \left\{\textbf{D}_l^{(t,i)}\right\}$ such as $\boldsymbol{\mathcal{D}}^t = \left[\boldsymbol{\mathcal{D}}^t_1,\cdots,\boldsymbol{\mathcal{D}}^t_L\right]$. We can clearly relate $t$ with $l$ as $l{=}t \text{ mod } L$ and $l{=}L$ when $l{=}0$. We can also define that the first loop of RA is when $t < L$.

\subsubsection{Parent Selection} The first step of the GA leverages on the current population, $\boldsymbol{\mathcal{D}}^t$ or $\boldsymbol{\mathcal{D}}^t_l$ for the centralized and sequential/parallel approaches respectively, to create an offspring population, defined as $\tilde{\boldsymbol{\mathcal{D}}}^t = \left\{\tilde{\textbf{D}}^{(t,i')}, i' = 1,\cdots,3N_{pop}\right\}$ and $\tilde{\boldsymbol{\mathcal{D}}}^t_l = \left\{\tilde{\textbf{D}}^{(t,i')}_l, i' = 1,\cdots,3N_{pop}\right\}$. The resulting populations have $3$ times the size of the former \cite{GA_Bento}. The first $2N_{pop}$ individuals are generated through the crossover and mutation operators while the last $N_{pop}$ ones are the original parent population, $\boldsymbol{\mathcal{D}}^t$ or $\boldsymbol{\mathcal{D}}^t_l$. To choose the $2N_{pop}$ parent population, we included a repeating selection scheme based on a tournament that chooses the best individual, $\textbf{D}^{(t,i'')}$ or $\textbf{D}^{(t,i'')}_l$, from a random set of $i''{=}1,\cdots,t_k$ elements. In this work, we prevent the same parent from being matched with itself.

\subsubsection{Crossover Operation} For the crossover operation, we defined a binary mask whose dimensions are $\textbf{b}^i \in \mathbb{B}^{K \times NL}$ and $\textbf{b}^i_l \in \mathbb{B}^{K \times N}$ for the centralized and sequential/parallel approaches respectively, along with a probability, $p_r$. The purpose of these binary masks is to combine the $2N_{pop}$ previously selected parent individuals and create two offspring solutions. The binary mask forces the first offspring to inherit from parent $1$ the genes where the mask is true and from parent $2$ the genes where the mask is false. The inverse procedure is applied to the second offspring. After this operation, the ensuing population is defined as $\hat{\boldsymbol{\mathcal{D}}}^t = \left\{\hat{\textbf{D}}^{(t,i')}\right\}$ or $\hat{\boldsymbol{\mathcal{D}}}^t_l = \left\{\hat{\textbf{D}}^{(t,i')}_l\right\}$. Assuming that $\textbf{D}^{(t,a)}$ and $\textbf{D}^{(t,b)}$ are the two selected parents for the centralized approach and  $\textbf{D}^{(t,a)}_l$ and $\textbf{D}^{(t,b)}_l$ are the two selected parents for the sequential and parallel approaches, the two offspring individuals are given as
\begin{equation} \label{eq:GAcrossover_centralized1}
   \hat{\textbf{D}}^{(t,i)} {=} \left\{
   \begin{array}{lcl}
   
     \textbf{D}^{(t,a)}, & \text{if } \textbf{b}^i > 0\\
     \textbf{D}^{(t,b)}, & \text{if } \textbf{b}^i \le 0, 
    \end{array}
    \right. \,
\end{equation}
\begin{equation} \label{eq:GAcrossover_centralized2}
   \hat{\textbf{D}}^{(t,\left(N_{pop}{+}i\right))} {=} \left\{
   \begin{array}{lcl}
   
     \textbf{D}^{(t,a)}, & \text{if } \textbf{b}^i > 0\\
     \textbf{D}^{(t,b)}, & \text{if } \textbf{b}^i \le 0,
    \end{array}
    \right. \,
\end{equation}  
and
\begin{equation} \label{eq:GAcrossover_sequential1}
   \hat{\textbf{D}}^{(t,i)}_l {=} \left\{
   \begin{array}{lcl}
   
     \textbf{D}^{(t,a)}_l, & \text{if } \textbf{b}^i_l > 0\\
     \textbf{D}^{(t,b)}_l, & \text{if } \textbf{b}^i_l \le 0, 
    \end{array}
    \right. \,
\end{equation}
 
\begin{equation} \label{eq:GAcrossover_sequential2}
   \hat{\textbf{D}}^{(t,\left(N_{pop}{+}i\right))}_l {=} \left\{
   \begin{array}{lcl}
   
     \textbf{D}^{(t,a)}_l, & \text{if } \textbf{b}^i_l > 0\\
     \textbf{D}^{(t,b)}_l, & \text{if } \textbf{b}^i_l \le 0. 
    \end{array}
    \right. \,
\end{equation}  

Additionally, the original population is included after the crossover by
 \begin{equation} \label{eq:GA_centralized3}
   \hat{\textbf{D}}^{(t,\left(2N_{pop}{+}i\right))} {=} \textbf{D}^{(t,i)},
\end{equation}
and
 \begin{equation} \label{eq:GA_sequential3}
   \hat{\textbf{D}}^{t\left(2N_{pop}{+}i\right)}_l {=} \textbf{D}^{ti}_l.
\end{equation}

\subsubsection{Mutation Operation} Following the crossover operation, the offspring populations, $\tilde{\boldsymbol{\mathcal{D}}}^t$ or $\tilde{\boldsymbol{\mathcal{D}}}^t_l$, are obtained by being subject to a mutation operation, which is performed with a probability of $p_m$. 

\begin{algorithm}[H]
\caption{GA algorithm}
\label{alg:ga}
\begin{algorithmic}[t]
\small
\STATE \textbf{Initialize:} $N_{pop}$, $t_k$, $N_m$, $N_{mi}$, $N_i$, $p_r$, $p_m$, 
 $p_{rm}$, $p_{mm}$, $p_{mi}$, $p_{ri}$ and $\varepsilon$

\STATE $t \gets 1$

\STATE $l \gets 1$

\STATE $p_r \gets p_{ri}$

\STATE $p_m \gets p_{mi}$

\STATE \textbf{case} $\mathbb{P}_1$: Initialize $\boldsymbol{\mathcal{D}}^t$ and $n_i \gets 0$
\STATE \quad \quad $\mathbb{P}_2$: Initialize $\boldsymbol{\mathcal{D}}^t_l$, $N_m \gets L \times N_m and$, and $n_i \gets 0$
\STATE \quad \quad $\mathbb{P}_3$: Initialize $\boldsymbol{\mathcal{D}}^t_l, l{=}1,\cdots,L$, and $\textbf{n}_i \gets \textbf{0}_{L} $

\WHILE{$\left(t < N_m \land n_i < N_i\right)$}

\IF{$\left(n_i \ge N_{mi}\right)$}
    \STATE $p_r \gets p_r + p_{rm}$
    \STATE $p_m \gets p_m + p_{mm}$
    \IF{$\left(p_r > p_{rmax}\right)$}
        \STATE $p_r \gets p_{rmax}$
    \ENDIF
    \IF{$\left(p_m > p_{mmax}\right)$}
        \STATE $p_m \gets p_{mmax}$
    \ENDIF
\ELSE 
    \STATE $p_r \gets p_{ri}$
    \STATE $p_m \gets p_{mi}$
\ENDIF  

\FOR{$i \gets 1$ to $N_{pop}$}

\STATE \textbf{case} $\mathbb{P}_1$: Calculate $f_1\left(\textbf{D}^{(t,i)}\right)$ according to Eq. \eqref{eq:MaxSum}
\STATE \quad \quad $\mathbb{P}_2$: Calculate $f_2\left(\textbf{D}^{(t,i)}_l\right)$ according to Eq. \eqref{eq:MaxSum_local}
\STATE \quad \quad $\mathbb{P}_3$: Calculate $f_3\left(\textbf{D}^{(t,i)}_l\right), l {=}1,\cdots,L$ according to Eq. \eqref{eq:MaxSum_parallel}

\ENDFOR

\STATE \textbf{case} $\mathbb{P}_1$: Calculate the best individual of $\boldsymbol{\mathcal{D}}^t, \textbf{D}^{(t,i_{best})}$
\STATE \quad \quad $\mathbb{P}_2$: Calculate the best individual of $\boldsymbol{\mathcal{D}}^t_l, \textbf{D}^{(t,i_{best})}_l$
\STATE \quad \quad $\mathbb{P}_3$: Calculate the best individuals of $\boldsymbol{\mathcal{D}}^t_l, \textbf{D}^{(t,i_{best})}_l, l {=} 1,\cdots,L$

\FOR{$i \gets 1$ to $2N_{pop}$}

\STATE \textbf{case} $\mathbb{P}_1$: Choose the best individual from the set $\textbf{D}^{(t,i'')}$ according to Eq. \eqref{eq:MaxSum} and $t_k$
\STATE \quad \quad $\mathbb{P}_2$: Choose the best individual from the set $\textbf{D}^{(t,i'')}_l$ according to Eq. \eqref{eq:MaxSum_local} and $t_k$
\STATE \quad \quad $\mathbb{P}_3$: Choose the best individuals from the sets $\textbf{D}^{(t,i'')}_l, l {=} 1,\cdots,L$ according to Eq. \eqref{eq:MaxSum_parallel} and $t_k$

\ENDFOR

\STATE Shuffle the parents in order to avoid degenerated combinations

\FOR{$i \gets 1$ to $N_{pop}$}

\STATE \textbf{case} $\mathbb{P}_1$: Define $\textbf{b}^i$ and calculate the $\hat{\boldsymbol{\mathcal{D}}}^t{=}\left\{\hat{\textbf{D}}^{(t,i')}\right\}$ population resulting from the crossover operation according to Eqs. \eqref{eq:GAcrossover_centralized1}, \eqref{eq:GAcrossover_centralized2} and \eqref{eq:GA_centralized3}

\STATE \quad \quad Calculate the $\tilde{\boldsymbol{\mathcal{D}}}^t{=}\left\{\tilde{\textbf{D}}^{(t,i')}\right\}$ population resulting from the mutation operation

\STATE \quad \quad $\mathbb{P}_2$: Define $\textbf{b}^i_l$ and calculate the $\hat{\boldsymbol{\mathcal{D}}}^t_l{=}\left\{\hat{\textbf{D}}^{(t,i')}_l\right\}$ population resulting from the crossover operation according to Eqs. \eqref{eq:GAcrossover_sequential1}, \eqref{eq:GAcrossover_sequential2} and \eqref{eq:GA_sequential3}

\STATE \quad \quad Calculate the $\tilde{\boldsymbol{\mathcal{D}}}^t_l{=}\left\{\tilde{\textbf{D}}^{(t,i')}_l\right\}$ population resulting from the mutation operation

\algstore{myalg}
\end{algorithmic}
\end{algorithm}

\begin{algorithm}                     
\begin{algorithmic}[t]        
\algrestore{myalg}
\small

\STATE \quad \quad $\mathbb{P}_3$: Define $\textbf{b}^i_l$ and calculate the $\hat{\boldsymbol{\mathcal{D}}}^t_l{=}\left\{\hat{\textbf{D}}^{(t,i')}_l\right\}$ population resulting from the crossover operation for $l{=}1,\cdots,L$ according to Eqs. \eqref{eq:GAcrossover_sequential1}, \eqref{eq:GAcrossover_sequential2} and \eqref{eq:GA_sequential3}

\STATE \quad \quad Calculate the $\tilde{\boldsymbol{\mathcal{D}}}^t_l{=}\left\{\tilde{\textbf{D}}^{(t,i')}_l\right\}, l{=}1,\cdots,L$ population resulting from the mutation operation

\ENDFOR

\STATE \textbf{case} $\mathbb{P}_1$: Calculate the best individual of $\tilde{\boldsymbol{\mathcal{D}}}^t, \tilde{\textbf{D}}^{(t,i_{best})}$
\STATE \quad \quad $\mathbb{P}_2$: Calculate the best individual of $\tilde{\boldsymbol{\mathcal{D}}}^t_l, \tilde{\textbf{D}}^{(t,i_{best})}_l$
\STATE \quad \quad $\mathbb{P}_3$: Calculate the best individual of $\tilde{\boldsymbol{\mathcal{D}}}^t_l, \tilde{\textbf{D}}^{(t,i_{best})}_l, l{=}1,\cdots,L$

\FOR{$i \gets 1$ to $N_{pop}-2$}

\STATE \textbf{case} $\mathbb{P}_1$: Choose the best individual from the set $\tilde{\textbf{D}}^{(t,i'')}$ according to Eq. \eqref{eq:MaxSum} and $t_k$ to generate $\textbf{D}^{(\left(t{+}1\right),i)}$
\STATE \quad \quad $\mathbb{P}_2$: Choose the best individual from the set $\tilde{\textbf{D}}^{(t,i'')}_l$ according to Eq. \eqref{eq:MaxSum_local} and $t_k$ to generate $\textbf{D}^{(\left(t{+}1\right),i)}_l$
\STATE \quad \quad $\mathbb{P}_3$: Choose the best individual from the set $\tilde{\textbf{D}}^{(t,i'')}_l$ according to Eq. \eqref{eq:MaxSum_parallel} and $t_k$ to generate $\textbf{D}^{(\left(t{+}1\right),i)}_l$ for $l{=}1,\cdots,L$

\ENDFOR

\STATE \textbf{case} $\mathbb{P}_1$: Complete $\boldsymbol{\mathcal{D}}^{t{+}1}$ by including $\textbf{D}^{(t,i_{best})}$ and $\tilde{\textbf{D}}^{(t,i_{best})}$

\STATE \quad \quad \textbf{if} $\left|f_1\left(\textbf{D}^{(t,i_{best})}\right) {-} f_1\left(\textbf{D}^{(t-1,i_{best})}\right)\right| {<} \varepsilon \Rightarrow n_i \gets n_i + 1$

\STATE \quad \quad \textbf{else} $\Rightarrow n_i \gets 0$

\STATE \quad \quad $\mathbb{P}_2$: Complete $\boldsymbol{\mathcal{D}}^{t{+}1}_l$ by including $\textbf{D}^{(t,i_{best})}_l$ and $\tilde{\textbf{D}}^{(t,i_{best})}_l$

\STATE \quad \quad \textbf{if} $\left|f_2\left(\textbf{D}^{(t,i_{best})}_l\right) - f_2\left(\textbf{D}^{(t-1,i_{best})}_l\right)\right| < \varepsilon \Rightarrow n_i \gets n_i + 1$

\STATE \quad \quad \textbf{else} $\Rightarrow n_i \gets 0$

\STATE \quad \quad $\mathbb{P}_3$: Complete $\boldsymbol{\mathcal{D}}^{t{+}1}_l$ by including $\textbf{D}^{(t,i_{best})}_l$ and $\tilde{\textbf{D}}^{(t,i_{best})}_l$ for $l{=}1,\cdots,L$

\STATE \quad \quad \textbf{if} $\left|f_3\left(\textbf{D}^{(t,i_{best})}_l\right) - f_3\left(\textbf{D}^{(t-1,i_{best})}_l\right)\right| < \varepsilon \Rightarrow \textbf{n}_i\left(l\right) \gets \textbf{n}_i\left(l\right) + 1, l{=}1,\cdots,L$

\STATE \quad \quad \textbf{else} $\Rightarrow \textbf{n}_i\left(l\right) \gets 0, l{=}1,\cdots,L$

\STATE $t \gets t + 1$

\STATE $l \gets t \text{ mod } L$

\IF{$\left(l = L\right)$}
    \STATE $l \gets L$
 \ENDIF

\ENDWHILE

\end{algorithmic}
\end{algorithm}

Since the AP-UE association matrix is binary, our mutation operation simply changes a gene to the opposite bit. We have also employed elitism in both optimization approaches in which we assure that the best individuals from $\boldsymbol{\mathcal{D}}^t$ or $\boldsymbol{\mathcal{D}}^t_l$ and $\tilde{\boldsymbol{\mathcal{D}}}^t$ or $\tilde{\boldsymbol{\mathcal{D}}}^t_l$ are included in the following generation $\boldsymbol{\mathcal{D}}^{t{+}1} = \left\{\textbf{D}^{(\left(t{+}1\right),i)}\right\}$ or $\boldsymbol{\mathcal{D}}^{t{+}1}_l = \left\{\textbf{D}^{(\left(t{+}1\right),i)}_l\right\}$. Furthermore, the remaining $N_{pop}{-}2$ solutions are selected through a random tournament with $t_k$. Finally, in order to maintain reach the global optimal solution, we increase the crossover and mutation probabilities by $p_{rm}$ and $p_{mm}$ if there is no improvement in the objective function for a given number of generations, $N_{mi}$, from their nominal values, $p_{ri}$ and $p_{mi}$ until they reach a maximum of $p_{rmax}$ and $p_{mmax}$.

\subsection{Computational Complexity and Fronthaul Signalling}

The GA implemented for centralized, sequential, and parallel approaches is summarized in Alg. \ref{alg:ga}. To determine the computational complexities of the CMRC, COSLP \cite{APS_Conceicao}, SMRC and PMRC approaches we must calculate the total number of arithmetic operations. We then present the computational complexity using the big-O notation.

The most complex operation is the calculation of the SE for each optimization problem. For the centralized approach, the total number of arithmetic operations to perform this operation can be approximated by 
\begin{equation} \label{eq:SE_MRC_C}
\resizebox{1\columnwidth}{!}{$
    O_{MRC}^{c}(f_1) =  3KLN^2 + 2K^2LN + KL^2N^2 + 2K^2L^2N + 2K^3LN,$}
\end{equation}
for the CMRC and 
\begin{equation} \label{eq:SE_OSLP_C}
\resizebox{1\columnwidth}{!}{$
    O_{OSLP}^{c}(f_1) = 5KLN^2 + 8K^2LN + 2KL^2N^2 + 2K^2L^2N + 2K^3LN + 3K^2L,$}
\end{equation}
for the COSLP approaches. For the SMRC, we have 
\begin{equation} \label{eq:SE_MRC_S}
\resizebox{1\columnwidth}{!}{$
\begin{aligned}
O_{MRC}^{s}(f_2) = \left\{\begin{array}{lcl} 
3KlN^2 + 2K^2lN + Kl^2N^2 + 2K^2l^2N + 2K^3lN, \text{ if } t < L, \\
3KLN^2 + 2K^2LN + KL^2N^2 + 2K^2L^2N + 2K^3LN, \text{otherwise}.
\end{array}
\right. \,
\\
\end{aligned}$}
\end{equation}
Also, according to Alg. \ref{alg:ga}  $O_{MRC}^{p}(f_3) {=}  O_{MRC}^{c}(f_1)$ from Eq. \eqref{eq:SE_MRC_C}. Therefore, the total number of arithmetic operations for the CMRC, PMRC, and COSLP approaches can be approximated by
\begin{equation} \label{eq:C_MRC_C}
\resizebox{1\columnwidth}{!}{$
    O_{MRC}^{c} = n_{iter}\left(4N_{pop}O_{MRC}^c(f_1) + 5N_{pop}t_{k}O_{MRC}^c(f_1) - 2t_{k}O_{MRC}^c(f_1)\right)$,}
\end{equation}
\begin{equation} \label{eq:P_MRC_C}
\resizebox{1\columnwidth}{!}{$
   O_{MRC}^{p} = n_{iter}\left(4N_{pop}O_{MRC}^p(f_3) + 5N_{pop}t_{k}O_{MRC}^p(f_3) - 2t_{k}O_{MRC}^p(f_3)\right)$,}
\end{equation}
and
\begin{equation} \label{eq:C_OSLP_C}
\resizebox{1\columnwidth}{!}{$
    O_{OSLP}^{c} = n_{iter}\left(4N_{pop}O_{OSLP}^c(f_1) + 5N_{pop}t_{k}O_{OSLP}^c(f_1) - 2t_{k}O_{OSLP}^c(f_1)\right)$,}
\end{equation}
where $n_{iter}$ is the total number of generations of the GA. The approximated number of arithmetic operations for the SMRC is 
\begin{equation} \label{eq:C_MRC_S}
\resizebox{1\columnwidth}{!}{$
    O_{MRC}^{s} = n_{iter}\left(4N_{pop}O_{MRC}^s(f_2) + 5N_{pop}t_{k}O_{MRC}^s(f_2) - 2t_{k}O_{MRC}^s(f_2)\right)$.}
\end{equation}
From previous Eqs \eqref{eq:SE_MRC_C}-\eqref{eq:C_MRC_S}, we can conclude that the computational complexity for all schemes follows $\mathcal{O}\left(K^3LN\right)$. However, the total number of operations is vastly different for each case. The COSLP approach requires much more arithmetic operations than the CMRC approach as can be seen by comparing Eqs. \eqref{eq:SE_MRC_C} and \eqref{eq:SE_OSLP_C} or, consequently, \eqref{eq:C_MRC_C}/\eqref{eq:P_MRC_C} and \eqref{eq:C_OSLP_C}. Furthermore, we can also conclude that the sequential approach for the MRC detection scheme has a significant reduction in the total computation complexity which can be corroborated by Eqs. \eqref{eq:SE_MRC_S} and \eqref{eq:C_MRC_S} as the first member of \eqref{eq:SE_MRC_S} is always smaller than Eq. \eqref{eq:SE_MRC_C}.

In \cite{RS_Shaik2}, the authors studied the fronthaul signalling capacity by measuring the total number of real symbols that each AP $l$ needs to share with the consecutive AP in one coherence block. The OSLP signal processing requires a total fronthaul signalling of $2K\left(\tau_c - \tau_p\right) + K^2$ real symbols since each AP $l$ has to forward the signal estimate $\hat{\textbf{s}}_l$ to the AP $l{+}1$ and the error covariance matrix, $\textbf{P}_l$, which contains the CE information. Similarly, the MRC signal processing requires a total fronthaul signalling of $2K\left(\tau_c - \tau_p\right) + K$ real symbols since each AP only needs the MMSE CE which is shared once in every coherence block.

In our analysis, the AP selection scheme requires additional information to be transmitted through the fronthaul link. Looking at Eq. \eqref{eq:SINR}, in order to calculate the SE, the APs need to share parts of the AP-UE association matrix, i.e. $\textbf{D}$. We must keep in mind that the CMRC or COSLP \cite{APS_Conceicao} approaches require all elements from $\textbf{D}$ to be transmitted to the CPU in order to efficiently perform the AP selection scheme whereas the SMRC AP selection approach is adapted to the sequential signal processing of the RS architecture \cite{RS_Shaik, RS_Shaik2} and the $l$th AP's LPU only requires the sub-matrix $\textbf{D}_{l-1}$ from previous AP to perform the AP selection optimization. Finally, the PMRC approach does not require any additional information to be transmitted between neighbor APs. Therefore, the total fronthaul signalling from the combination of the AP selection and equalization schemes is given by

\begin{equation} \label{eq:F_MRC_C}
    F_{MRC}^{c} = 2K\left(\tau_c - \tau_p\right) + K + KNL,
\end{equation}

\begin{equation} \label{eq:F_OSLP_C}
    F_{OSLP}^{c} = 2K\left(\tau_c - \tau_p\right) + K^2 + KNL,
\end{equation}

\begin{equation} \label{eq:F_MRC_S}
    F_{MRC}^{s} = 2K\left(\tau_c - \tau_p\right) + K + KN,
\end{equation}

\begin{equation} \label{eq:F_MRC_P}
    F_{MRC}^{p} = 2K\left(\tau_c - \tau_p\right) + K.
\end{equation}

Clearly, $F_{MRC}^{p} < F_{MRC}^s < F_{MRC}^c < F_{OSLP}^c$, which indicates that the parallel approach does not increase the latency of RS network, while the sequential AP selection approach is capable of reducing the fronthaul signalling load when compared to the centralized one.

The GA MH algorithmic approach is capable of operating at the large-scale fading time scales. This facilitates the operation at either the CPU (for the centralized schemes) or the LPU (for the sequential and parallel schemes) before the RS UL communication occurs. Considering the SE expressions involved, which depend on various factors such as the AP selection allocation matrix and the CE vectors, it is crucial to assume that the CE vectors remain valid for a specified period — true for a given coherence time-frequency block. Consequently, the output solutions from the GA approach are valid for that coherence block. This implies that the AP selection matrix is updated at the large-scale time scale, and the suitability of the algorithm for (near) real-time applications is evident. The algorithm can effectively adapt to the network environment, requiring updates only once per coherent block.

\section{Simulation Results}

The RS network is implemented in a general indoor scenario with a squared shape of $D \times D$ m$^2$, whose UEs are randomly placed in the far-field of the AP antennas region given by the Fraunhofer distance, $D_{fd}{=} \frac{N^2\lambda}{2}$ where $\lambda$ is the wavelength. The APs and UEs are placed at a height of $h_{AP}$ and $h_{UE}$. The large-scale fading coefficient is defined by
\begin{equation} \label{eq:LargeScaleModel}
\resizebox{1\columnwidth}{!}{$
   \beta_{kl}\left(d_{kl}\right) = -36.7\log_{10}\left(\sqrt{d_{kl}^2}\right) -22.7 - 26\log_{10}\left(f_c\right) + F_{kl}\left(d_{kl} > d_{sh}\right),$}
\end{equation}
where $d_{kl}$ is distance between UE $k$ and AP $l$, $f_c$ is the carrier frequency in GHz, $d_{sh}$ is the decorrelation distance and $F_{kl} \sim \mathcal{N}\left(0,\sigma_{sh}^2\right)$ is the shadow fading term following
\begin{equation} \label{eq:ShadowFading}
    \mathbb{E}\left\{F_{kl}F_{ij}\right\} = \sigma_{sh}^2 \left(\Delta 2^{-\frac{\nu_{ki}}{d_c}} + \left(1 - \Delta\right) 2^{-\frac{\zeta_{lj}}{d_c}}\right),
\end{equation}
where it takes into consideration both contributions, $\Delta$, from neighbor UEs and APs through $\nu_{ki}$ and $\zeta_{lj}$. The UL bandwidth is $B$ MHz and the noise power in dBm is $\sigma^2{=} 10\text{log}_{10}\left(k_BT_0\right) + 10\text{log}_{10}\left(B\right) + \sigma_f^2$, where $k_B$ is the Boltzmann constant in J/K, $T_0$ is the noise temperature in K and $\sigma_f^2$ is the noise figure in dB. The antenna spacing is $d_H$ and the multiple rays reach an AP according to angular Gaussian distribution with standard deviation  $\sigma_{\phi}$ \cite{MMSE_Bjornson}. Since we are considering a MH algorithm, the results of the GA-based APS scheme were obtained through $N_a{=}10$ independent realizations. The GA and RS network parameters are shown in Tables \ref{table_GA} and \ref{table_RS}, respectively. 

\begin{table}[t]%
\centering
\caption{GA Parameters}
\label{table_GA}
\begin{tabular}[t]{|c c | c c | c c |}
\hline
  \textbf{Parameter} & \textbf{Value} & \textbf{Parameter} & \textbf{Value} & \textbf{Parameter} & \textbf{Value} \\
\hline
  $N_{pop}$ & $100$ & $t_k$ & $2$ & $p_{mi}$ & $0.01$ \\
\hline
  $N_m$ & $1000$ & $p_{ri}$ & $0.75$ & $N_{mi}$ & $5$\\
\hline
  $N_i$ & $100$ & $p_{mi}$ & $0.01$ & $p_{rmax}$ & $1$\\
\hline
  $\varepsilon$ & $10^{-5}$ & $p_{rm}$ & $0.1$ & $p_{mmax}$ & $0.1$\\
\hline
\end{tabular}
\end{table}

\begin{table}[t]%
\centering
\caption{RS Parameters}
\label{table_RS}
\resizebox{1\columnwidth}{!}{
\begin{tabular}[t]{|c c | c c | c c |}
\hline
  \textbf{Parameter} & \textbf{Value} & \textbf{Parameter} & \textbf{Value} & \textbf{Parameter} & \textbf{Value} \\
\hline
  $D$ & $20$ m & $f_c$ & $2$ GHz & $\Delta$ & $0.7$ \\
\hline
  $L$ & $\left\{8,10\right\}$ & $h_{AP}$ & $6$ m & $d_c$ & $9$ m \\
\hline
  $N$ & $\left\{2,4\right\}$ & $h_{UE}$ & $1$ m & $d_{sh}$ & $13$ m\\
\hline
  $K$ & $\left\{5,10\right\}$ & $D_{fd}$ & $\left\{0.3,1.2\right\}$ m & $d_H$ & $\frac{\lambda}{2}$ \\
\hline
  $\tau_c$ & $300$ & $B$ & $250$ MHz & $\sigma_{\phi}$ & $7.5$ º\\
\hline
  $\tau_p$ & $\left\{4,8\right\}$ & $\sigma^2_f$ & $4$ dB & $k_B$ & $1.38\times 10^{-23}$ J/K \\
\hline
  $P_{max}$ & $30$ mW & $\sigma_{sh}^2$ & $2$ dB  & $T_0$ & $290$ K \\
\hline
\end{tabular}}
\end{table}

\subsection{Centralized vs Sequential vs Parallel AP Selection}

In the following simulations, we will evaluate the performance of the CMRC and COSLP AP selection approaches \cite{APS_Conceicao} and the SMRC, PMRC AP selection schemes in four scenarios. These scenarios contain different levels of UE-AP densities.

\begin{itemize}
    \item Scenario 1: $K{=}10$, $N{=}4$ and $L{=}12$.
    \item Scenario 2: $K{=}5$, $N{=}2$ and $L{=}8$.
    \item Scenario 3: $K{=}5$, $N{=}4$ and $L{=}12$.
    \item Scenario 4: $K{=}10$, $N{=}2$ and $L{=}8$.
\end{itemize}

\begin{figure}[t!]
\centering
\begin{minipage}{0.5\columnwidth}
\centering
\subfigure[]{\includegraphics[width=1\columnwidth]{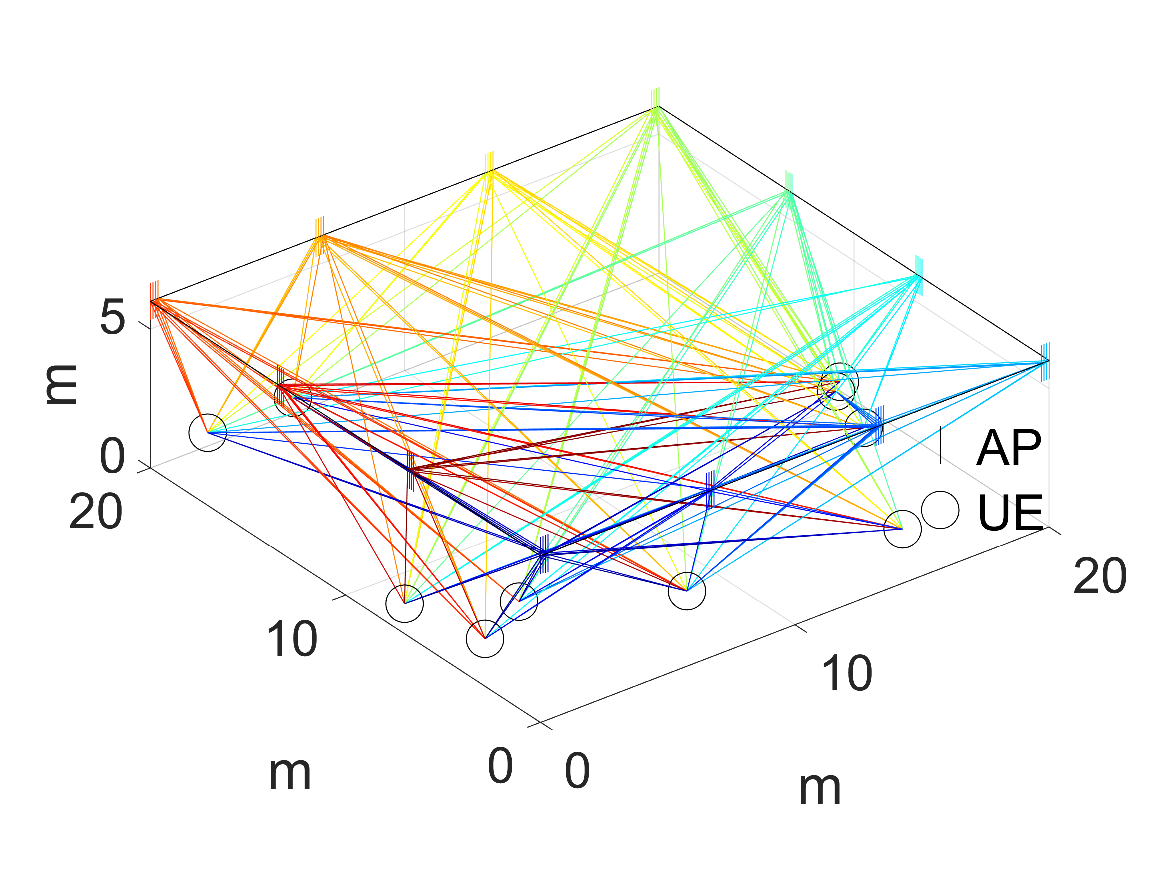}}
\end{minipage}%
\begin{minipage}{0.5\columnwidth}
\centering
\subfigure[]{\includegraphics[width=1\columnwidth]{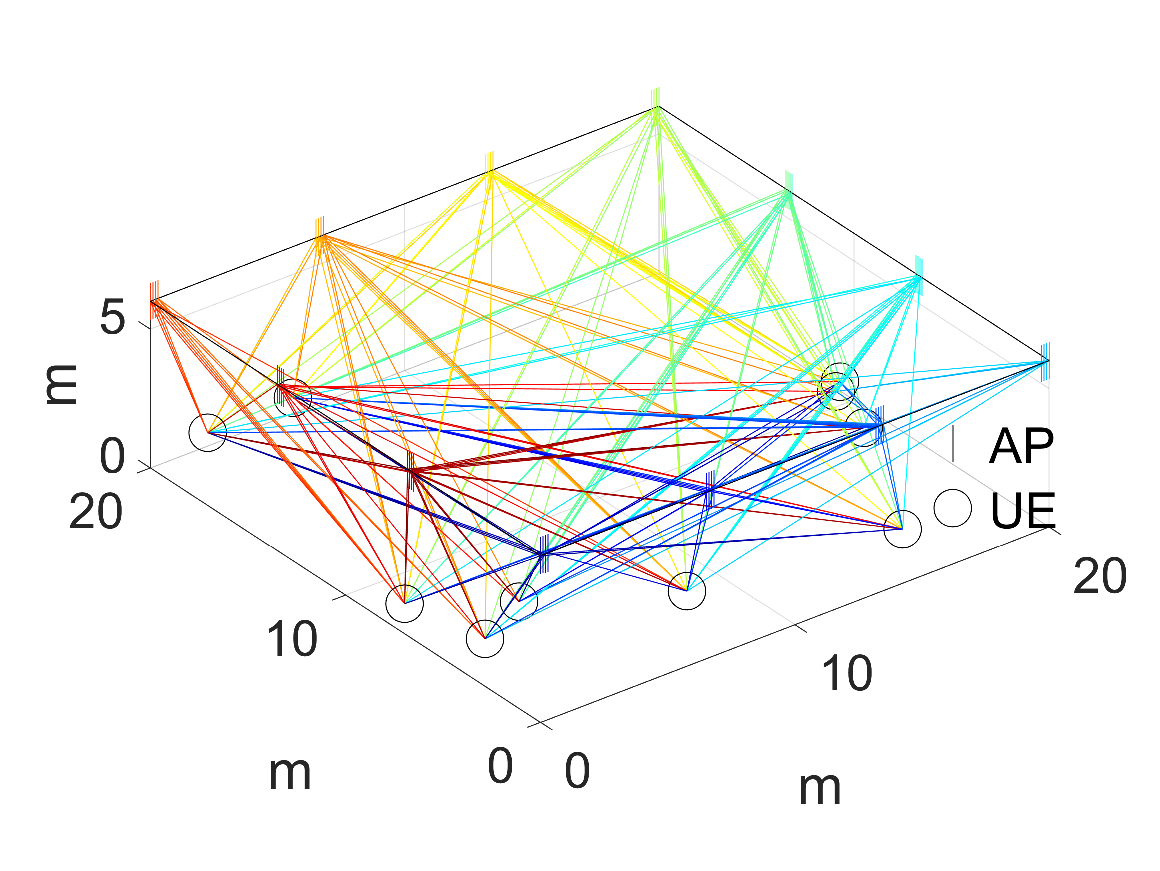}} \\
\end{minipage}%
\\
\begin{minipage}{0.5\columnwidth}
\centering
\subfigure[]{\includegraphics[width=1\columnwidth]{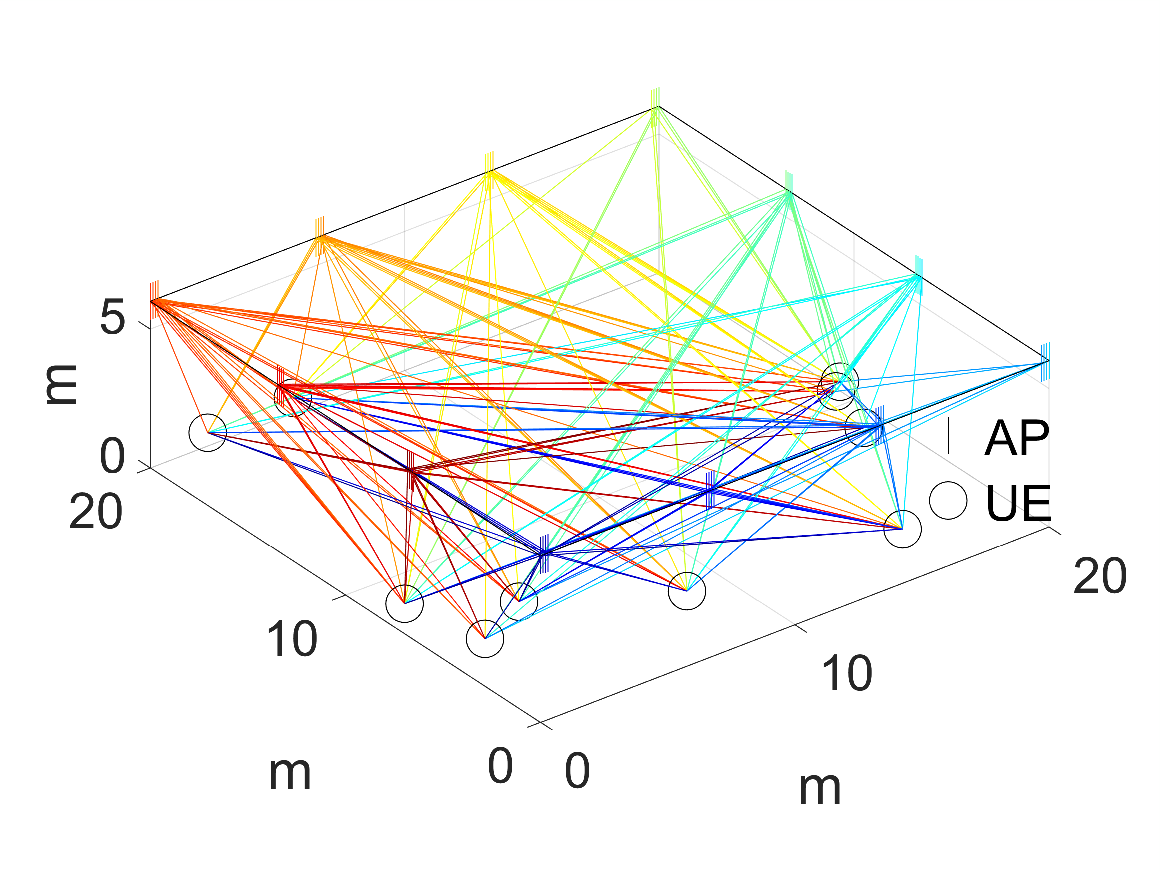}}
\end{minipage}%
\begin{minipage}{0.5\columnwidth}
\centering
\subfigure[]{\includegraphics[width=1\columnwidth]{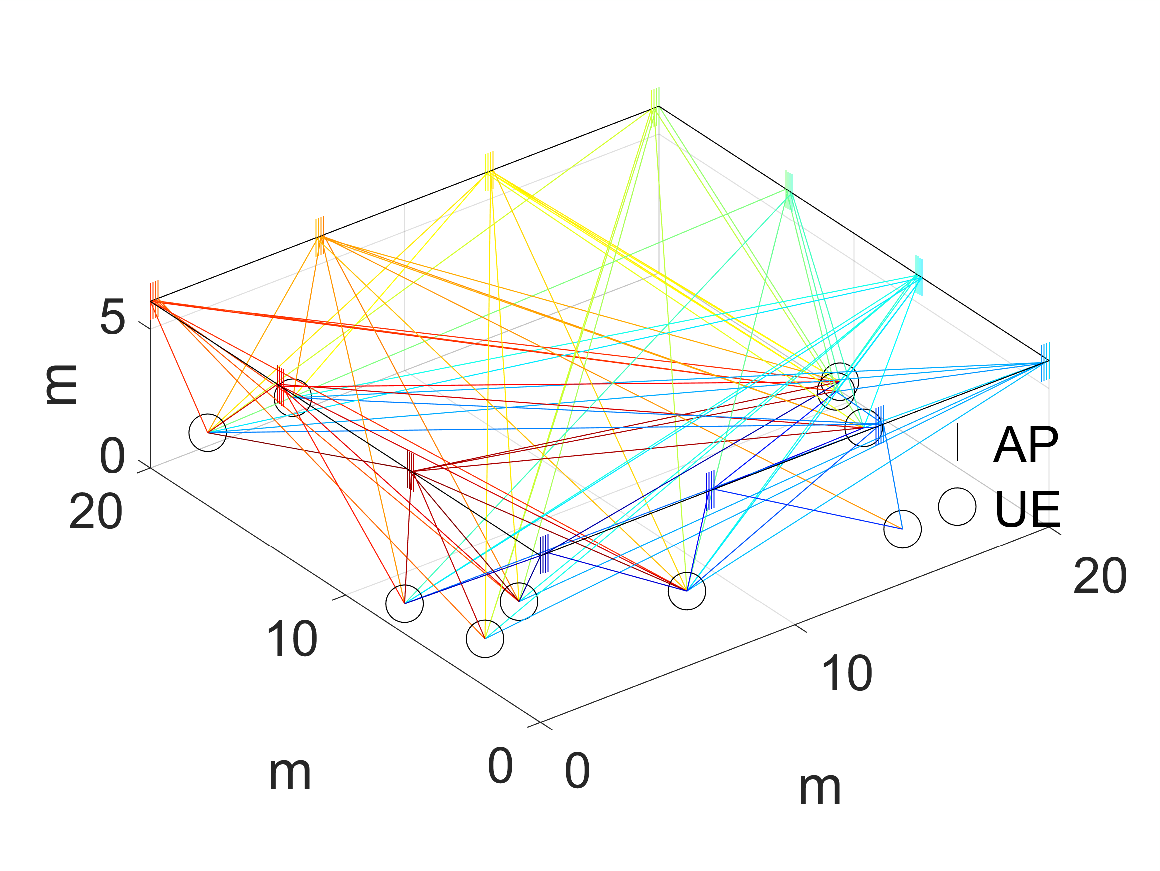}}
\end{minipage}%
\caption{Optimized AP-UE association for scenario 1 with (a) CMRC \cite{APS_Conceicao} (b) COSLP \cite{APS_Conceicao}, (c) SMRC and (d) PMRC}.
\label{fig:APS_scenario1}
\end{figure}

\begin{figure}[t!]
\centering
\begin{minipage}{0.5\columnwidth}
\centering
\subfigure[]{\includegraphics[width=1\columnwidth]{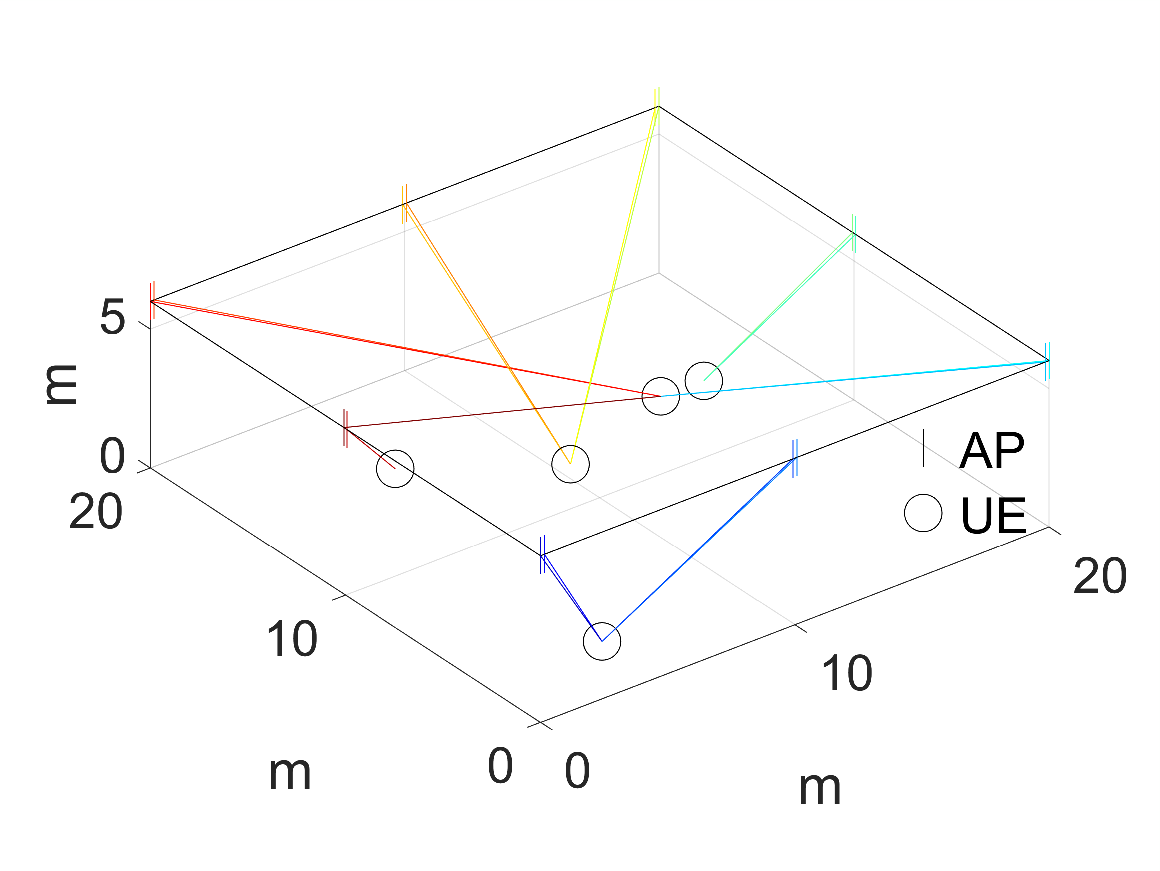}}
\end{minipage}%
\begin{minipage}{0.5\columnwidth}
\centering
\subfigure[]{\includegraphics[width=1\columnwidth]{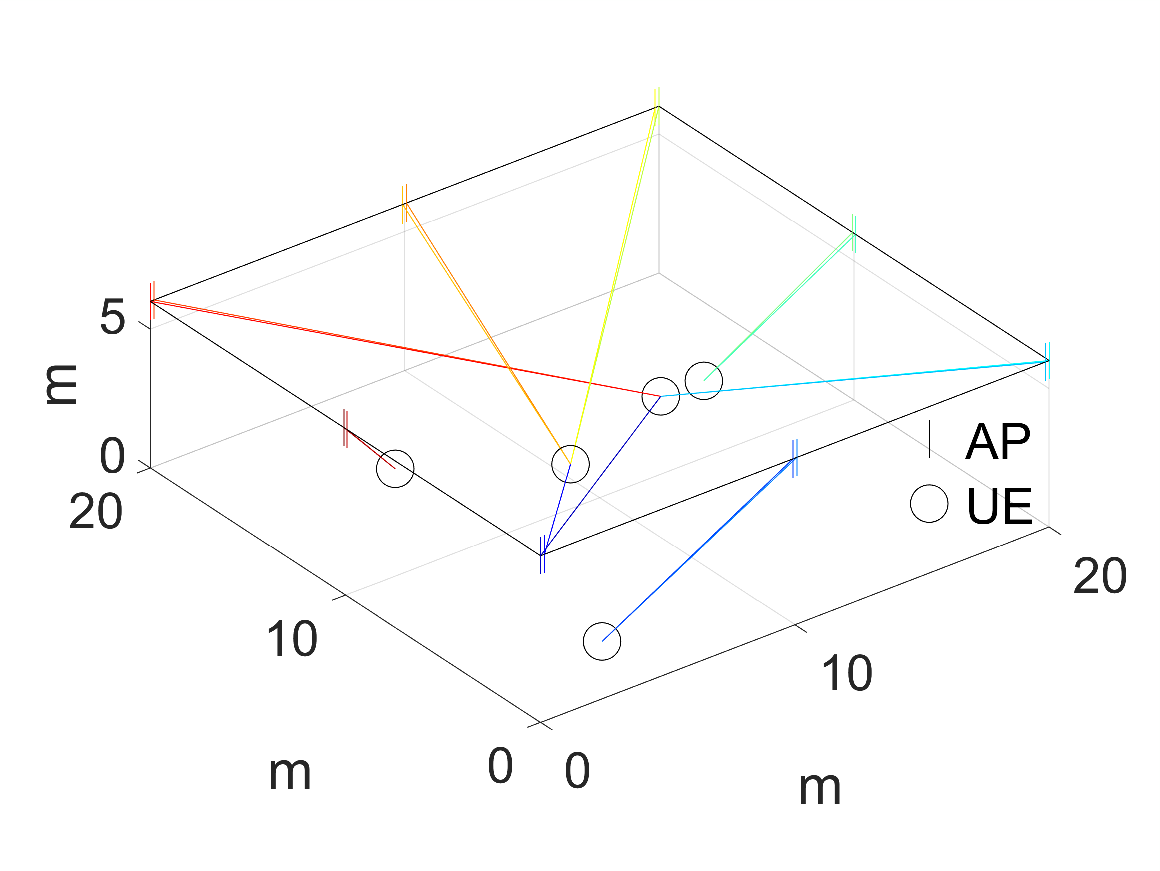}} \\
\end{minipage}%
\\
\begin{minipage}{0.5\columnwidth}
\centering
\subfigure[]{\includegraphics[width=1\columnwidth]{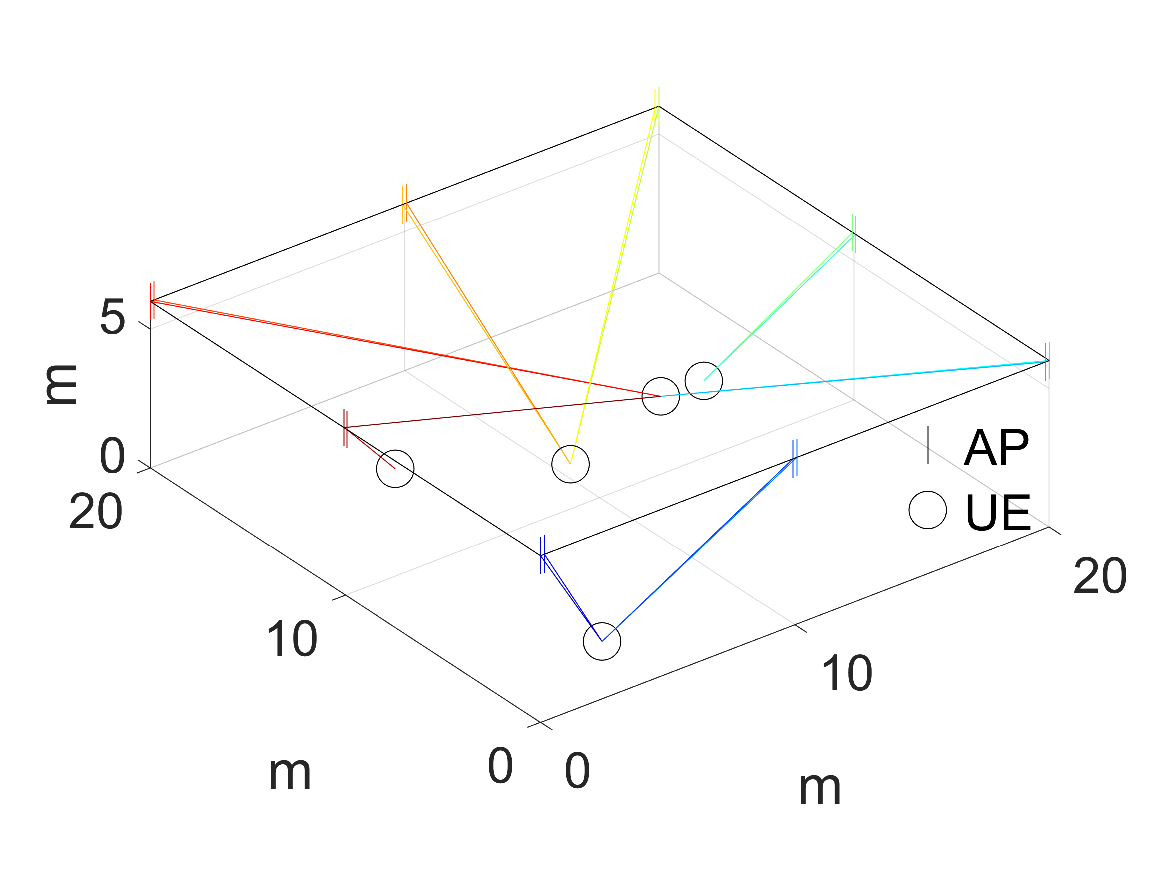}}
\end{minipage}%
\begin{minipage}{0.5\columnwidth}
\centering
\subfigure[]{\includegraphics[width=1\columnwidth]{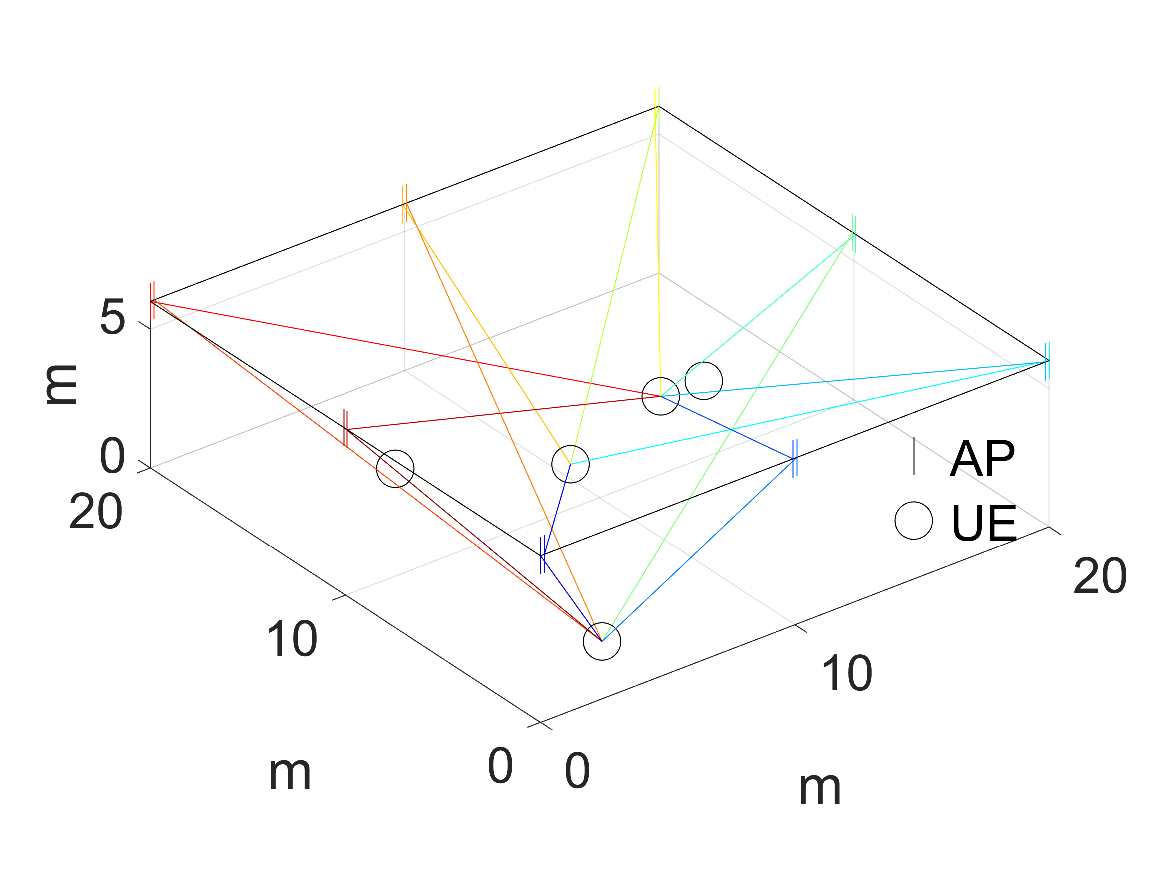}}
\end{minipage}%
\caption{Optimized AP-UE association for scenario 2 with (a) CMRC \cite{APS_Conceicao}, (b) COSLP \cite{APS_Conceicao}, (c) SMRC and (d) PMRC}.
\label{fig:APS_scenario2}
\end{figure}

\begin{figure}[t!]
\centering
\begin{minipage}{0.5\columnwidth}
\centering
\subfigure[]{\includegraphics[width=1\columnwidth]{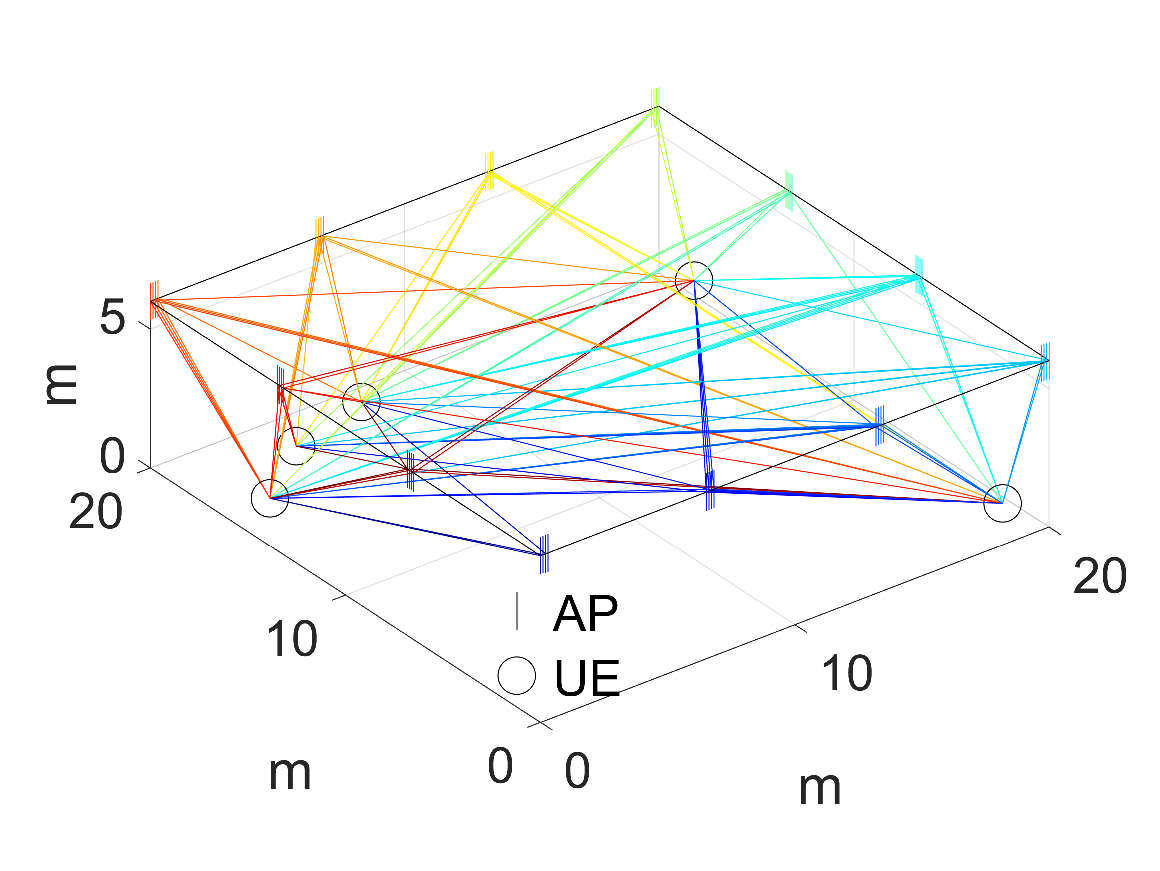}}
\end{minipage}%
\begin{minipage}{0.5\columnwidth}
\centering
\subfigure[]{\includegraphics[width=1\columnwidth]{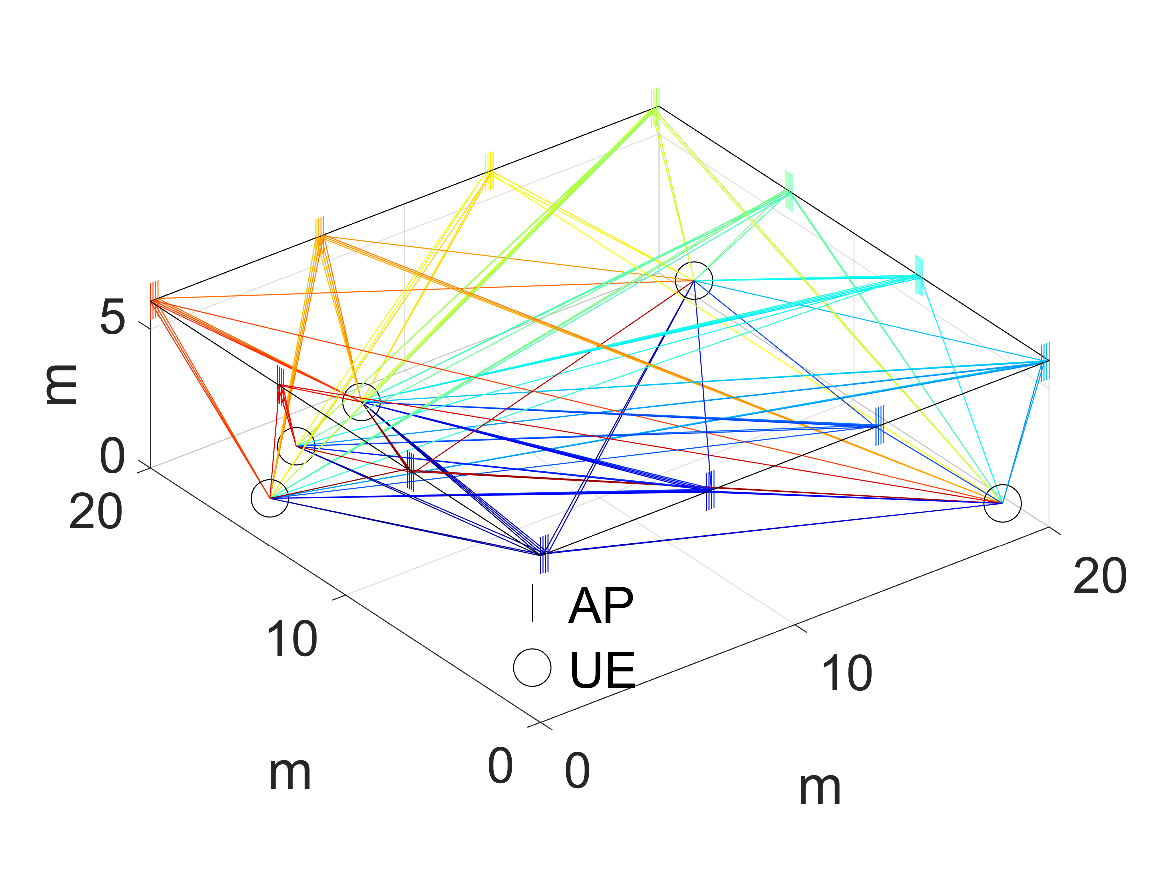}} \\
\end{minipage}%
\\
\begin{minipage}{0.5\columnwidth}
\centering
\subfigure[]{\includegraphics[width=1\columnwidth]{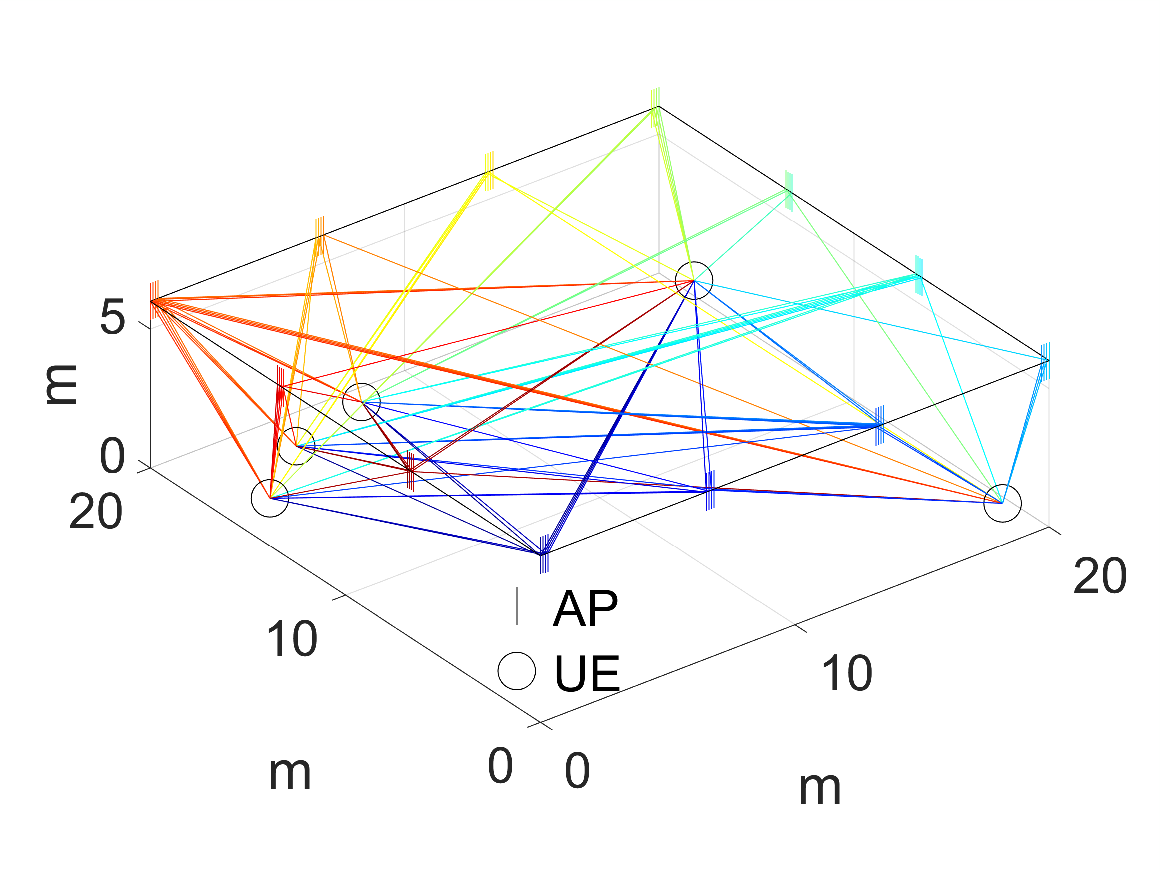}}
\end{minipage}%
\begin{minipage}{0.5\columnwidth}
\centering
\subfigure[]{\includegraphics[width=1\columnwidth]{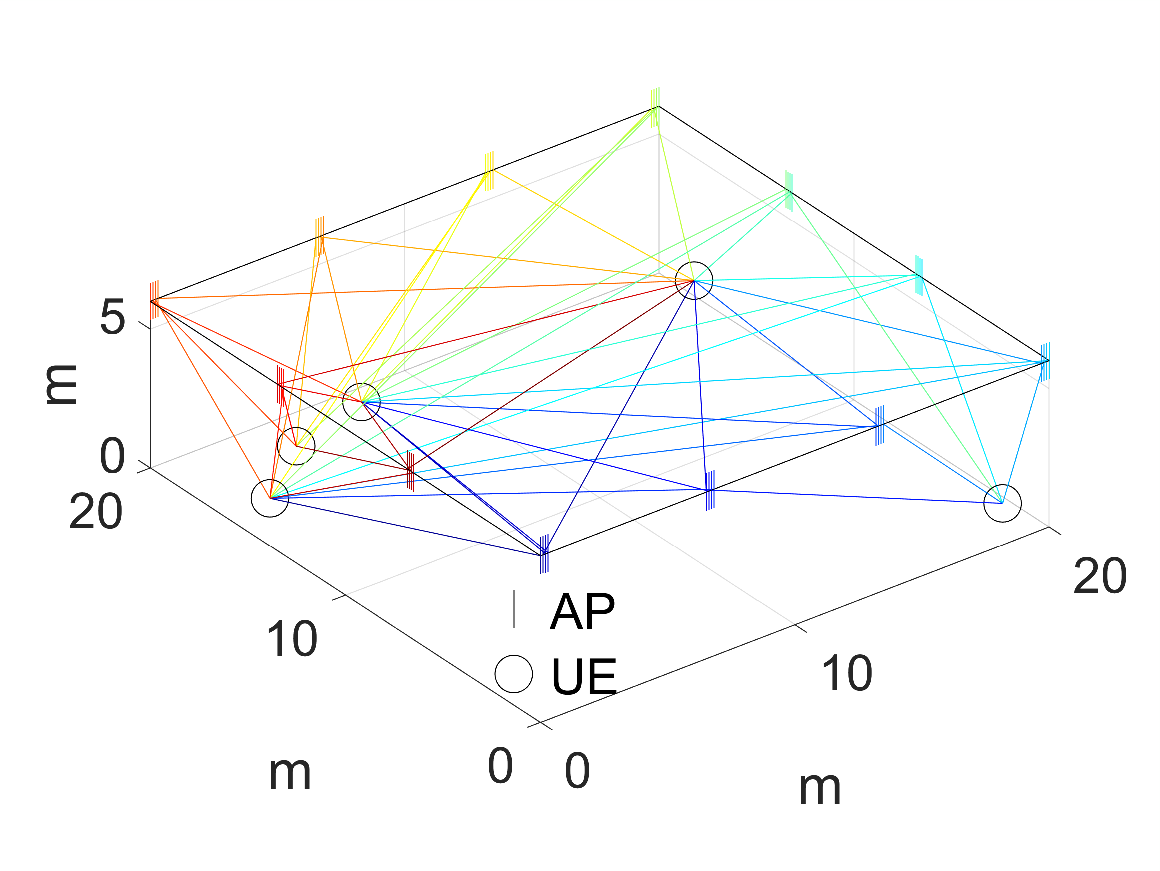}}
\end{minipage}%
\caption{Optimized AP-UE association for scenario 3 with (a) CMRC \cite{APS_Conceicao}, (b) COSLP \cite{APS_Conceicao}, (c) SMRC and (d) PMRC}.
\label{fig:APS_scenario3}
\end{figure}

\begin{figure}[t!]
\centering
\begin{minipage}{0.5\columnwidth}
\centering
\subfigure[]{\includegraphics[width=1\columnwidth]{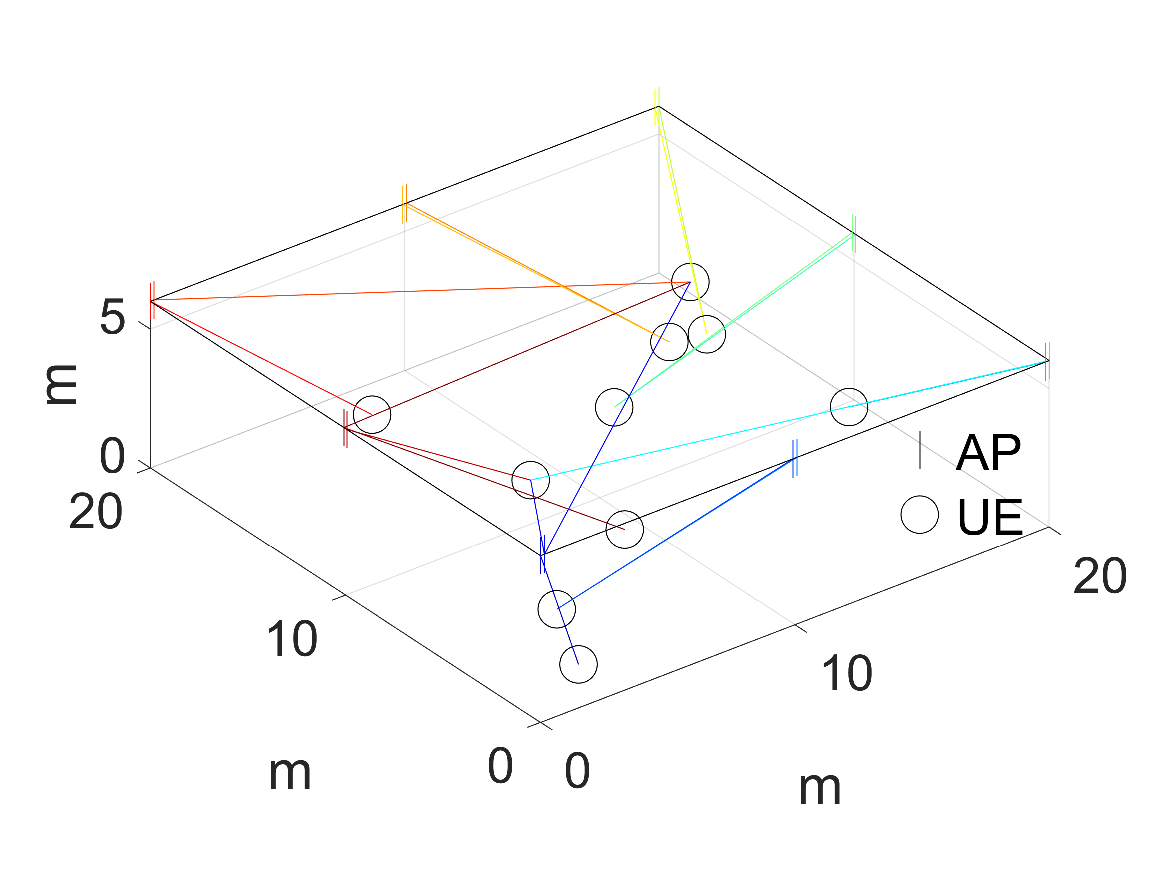}}
\end{minipage}%
\begin{minipage}{0.5\columnwidth}
\centering
\subfigure[]{\includegraphics[width=1\columnwidth]{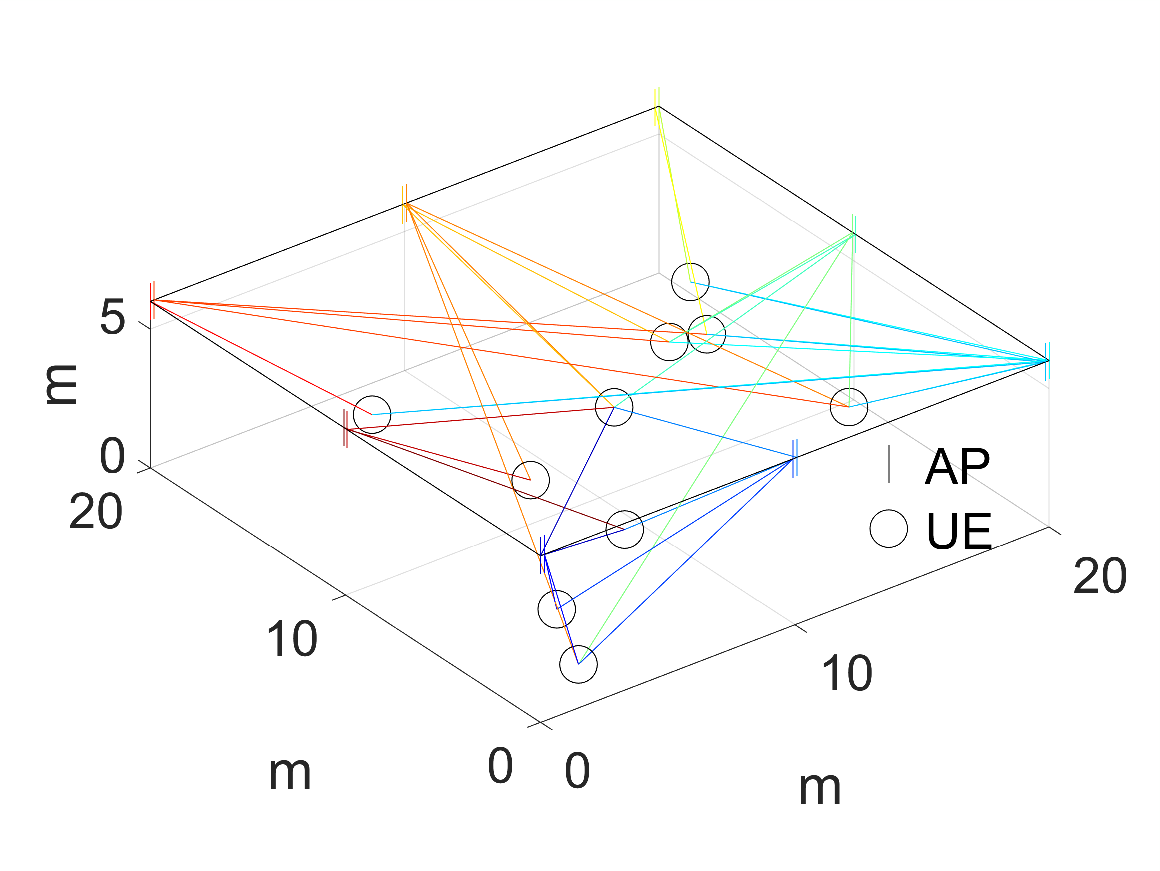}} \\
\end{minipage}%
\\
\begin{minipage}{0.5\columnwidth}
\centering
\subfigure[]{\includegraphics[width=1\columnwidth]{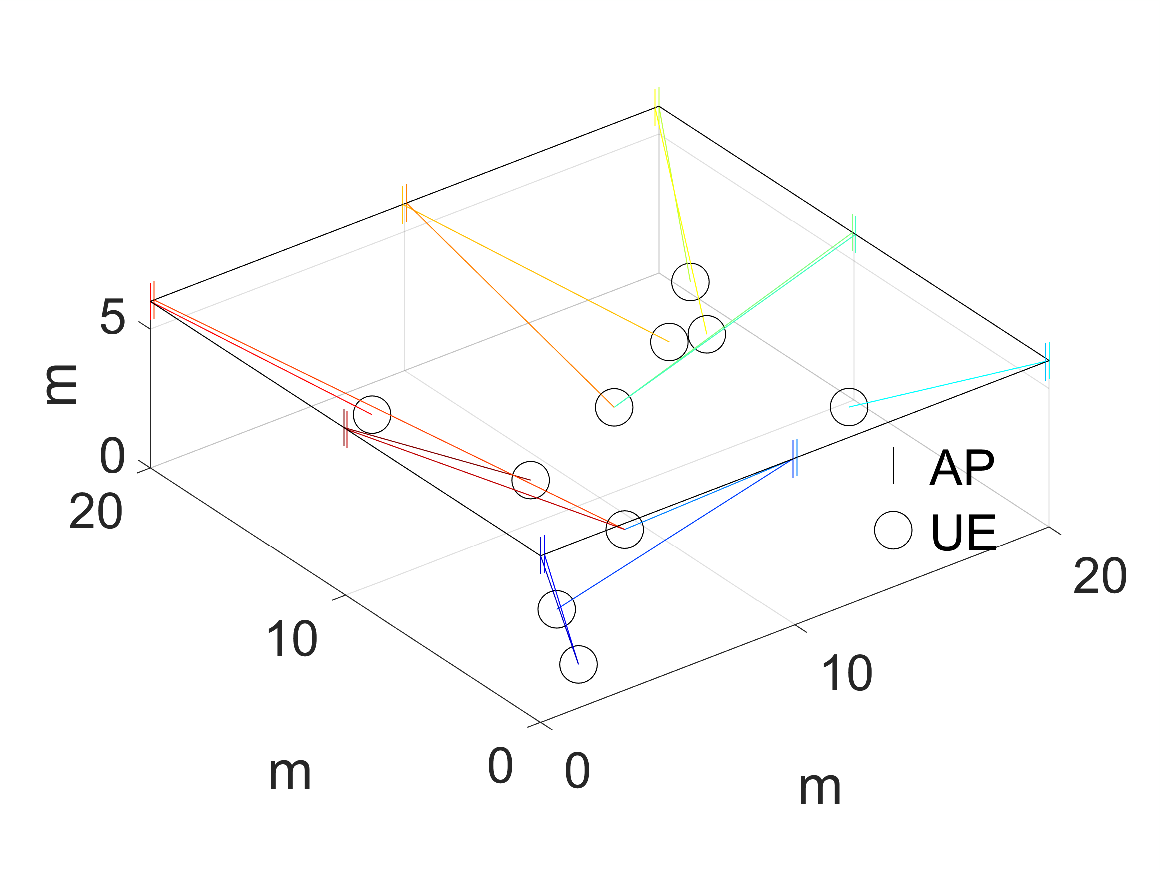}}
\end{minipage}%
\begin{minipage}{0.5\columnwidth}
\centering
\subfigure[]{\includegraphics[width=1\columnwidth]{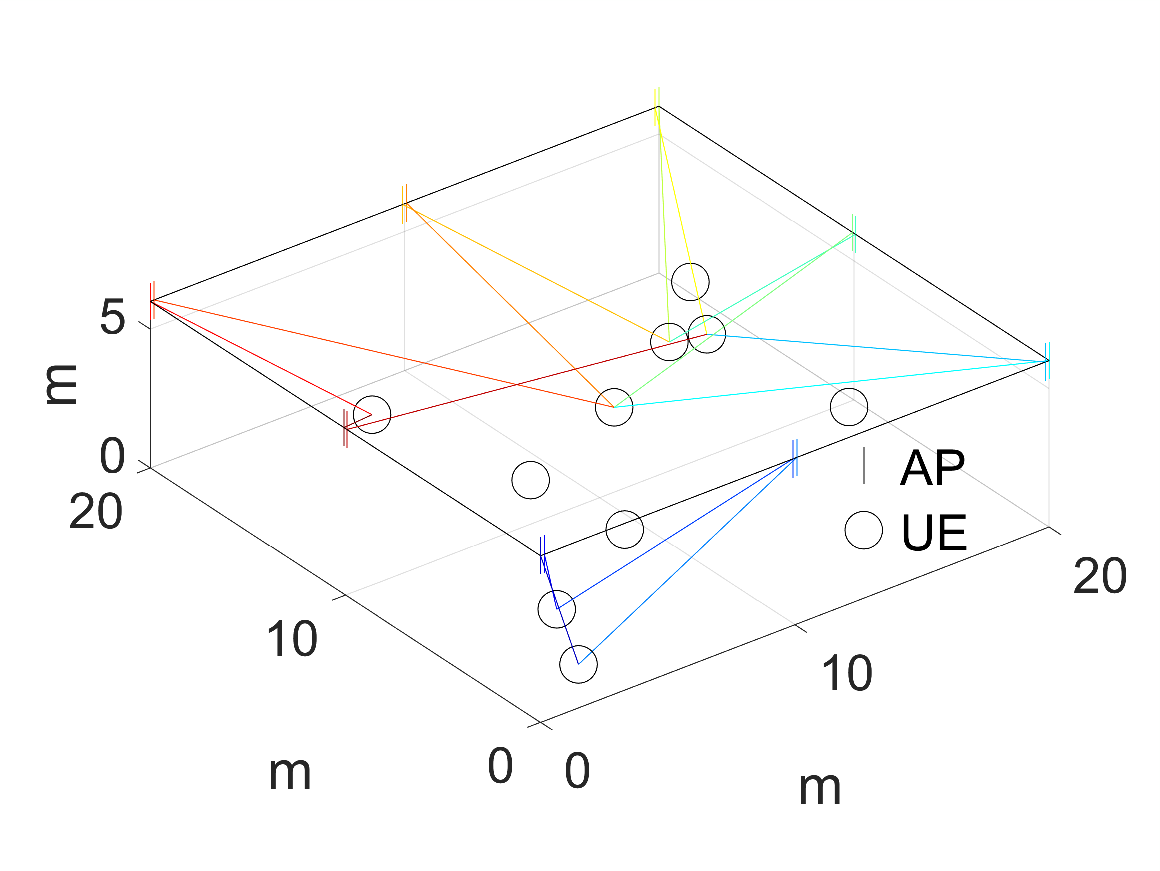}}
\end{minipage}%
\caption{Optimized AP-UE association for scenario 4 with (a) CMRC \cite{APS_Conceicao}, (b) COSLP \cite{APS_Conceicao}, (c) SMRC and (d) PMRC}.
\label{fig:APS_scenario4}
\end{figure}

Figs. \ref{fig:APS_scenario1}-\ref{fig:APS_scenario4} present a graphical representation of the AP-UE association resulting from the best realization of the AP selection scheme obtained by the GA for each implementation and scenario.

The complexity of the AP selection optimization problem can also be inferred by observing \ref{fig:APS_scenario1}-\ref{fig:APS_scenario4} as it can be related to the number of RAs that are performed between the UEs and the APs' antennas. Due to this, scenarios 1 and 3 contain are much more computationally complex as the total number of antennas in the RS network is $48$ in contrast to the $16$ from scenarios 2 and 4.

\begin{figure}[t!]
\centering
\subfigure[]{\includegraphics[width=1\columnwidth]{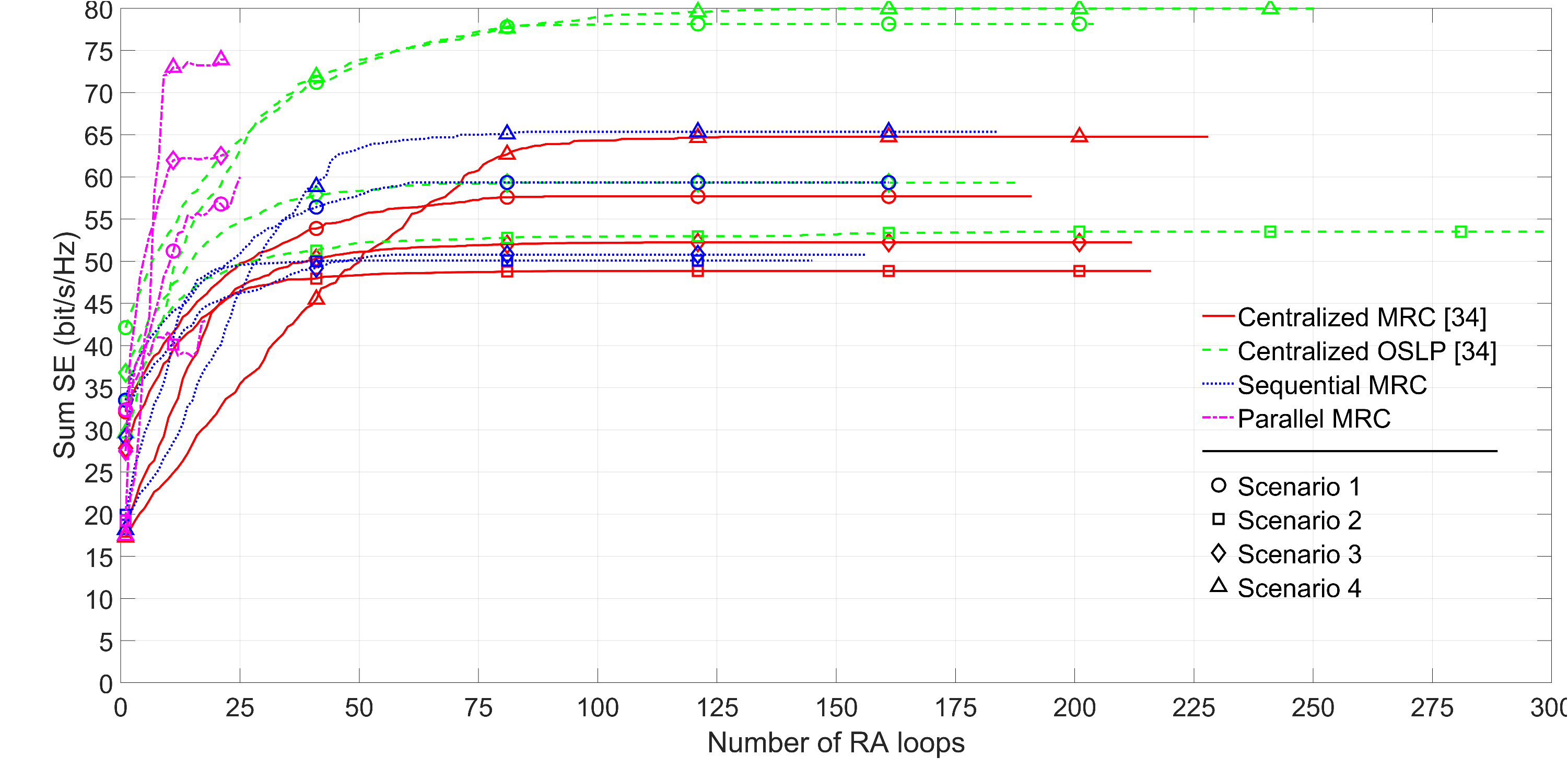}}
\hfill
\subfigure[]{\includegraphics[width=1\columnwidth]{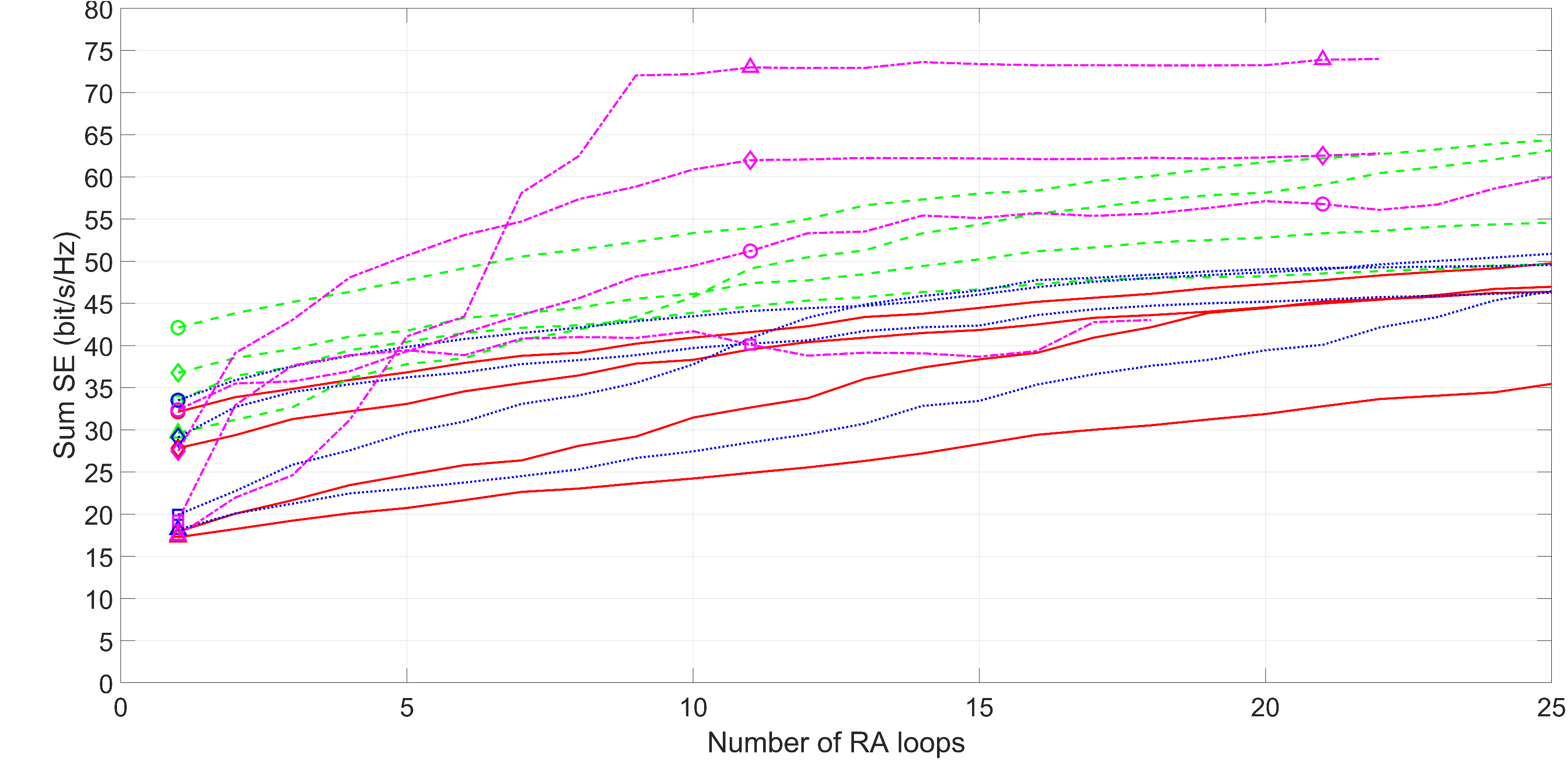}}
\caption{Sum SE, in bits/s/Hz, obtained with CMRC, COSLP \cite{APS_Conceicao}, SMRC, and PMRC for all scenarios: (a) full range and (b) zoomed in the 0-25 RA range.}
\label{fig:evolution_schemes}
\end{figure}

Fig. \ref{fig:evolution_schemes} depicts the performance of the GA-based AP selection algorithm with the CMRC, COSLP, SMRC, and PMRC approaches for scenarios 1 through 4. For this, we plot the average of the sum SE (in bits/s/Hz), which is obtained from the $N_a$ realizations, that each equalization method can obtain as a function of the number of RA loops.

As can be seen from Fig. \ref{fig:evolution_schemes}, the SMRC scheme is able to converge faster to reasonably similar values of total throughput when compared to CMRC in all scenarios, which also reflects on a similar AP-UE association matrix. Additionally, the PMRC can converge even faster than the previously mentioned methods, while outperforming them in scenarios 3 and 4. However, it can also lead to a worse solution, as observed in scenario 2. Furthermore, as can be expected, the COSLP scheme provides a better SE performance when compared to both MRC approaches, at the expense of increased computational complexity.

For scenario 1, at $10$ RA loops, the COSLP reached a sum SE of $53.3$ bits/s/Hz. The CMRC, SMRC, and PMRC display a sum SE of $40.9$, $43.5$, and $49.5$ bits/s/Hz respectively. At $25$ RA loops, the GA employing the PMRC method converges, leading to a sum SE of $60.0$ bits/s/Hz. At $50$ RA loops, the COSLP is able to attain $73.4$ bits/s/Hz of sum SE, while the CMRC and SMRC reach a sum SE of $55.4$ and $57.9$ bits/s/Hz, respectively. The CMRC converges at around $91$ RA loops, with the final solution exhibiting a sum SE of $57.7$ bits/s/Hz. At around $61$ RA loops, the SMRC approach converges to the final solution, obtaining a sum SE of $59.3$ bits/s/Hz. The COSLP has a sum SE of $78.1$ bits/s/Hz, requiring around $104$ RA loops to converge.

For scenario 2, at $10$ RA loops, the COSLP can obtain a sum SE of $43.9$ bits/s/Hz, while the CMRC, SMRC, and PMRC are able to obtain a sum SE of $31.5$, $37.8$, and $41.7$ bits/s/Hz respectively. The PMRC has a final sum SE of $43$ bits/s/Hz and requires $18$ RA loops. At around $45$ RA loops, the SMRC approach converges to the final solution, obtaining a sum SE of $50.0$ bits/s/Hz, while the CMRC and COSLP reach a sum SE of $48.2$ bits/s/Hz and $51.6$ bits/s/Hz, respectively. The CMRC converges at around $116$ RA loops with a total SE of $48.9$ bits/s/Hz, while the COSLP has $53.0$ bits/s/Hz. The latter requires a total of $200$ RA loops to converge the sum SE to $53.5$ bits/s/Hz.

For scenario 3, at $10$ RA loops, the COSLP has a performance of $46.1$ bits/s/Hz, while the CMRC, SMRC, and PMRC schemes contain a sum SE of $38.3$, $39.7$, and $60.9$ bits/s/Hz, respectively. The PMRC has a final sum SE of $62.8$ bits/s/Hz and requires $22$ RA loops. At $50$ RA loops, the COSLP is capable of reaching a total of $58.7$ bits/s/Hz of sum SE, while the CMRC and SMRC can optimize the sum SE to $51.1$ and $50.3$ bits/s/Hz, respectively. At around $112$ and $56$ RA loops, both CMRC and SMRC approaches converge to the final solution, obtaining a sum SE of $52.2$ and $50.7$ bits/s/Hz, respectively. The COSLP requires around $89$ RA loops to converge to a total of $59.3$ bits/s/Hz.

For scenario 4, at $10$ RA loops, the COSLP is optimized to provide a sum SE of $45.8$ bits/s/Hz, while the CMRC, SMRC, and PMRC optimization output a sum SE of $24.2$, $27.4$, and $72.2$ bits/s/Hz, respectively. At $22$ RA loops, the GA employing the PMRC method converges, leading to a sum SE of $74$ bits/s/Hz. At $50$ RA loops, the COSLP can reach $73.9$ bits/s/Hz of sum SE, while the CMRC and SMRC can provide a sum SE of $49.9$ bits/s/Hz and $63.4$ bits/s/Hz, respectively. The performance gap between CMRC and SMRC approaches is significant as the SMRC scheme converges to the final sum SE of $65.3$ bits/s/Hz at around $84$ RA loops. At the same time, its centralized counterpart only acquires a sum SE of $63.4$ bits/s/Hz and the COSLP has $78.0$ bits/s/Hz of sum SE. The CMRC requires around $128$ RA loops to converge to $64.7$ bits/s/Hz of sum SE while the COSLP has $79.7$ bits/s/Hz of sum SE. This approach requires around $151$ RA loops to show a sum SE of $80$ bits/s/Hz.

This whole analysis also showed that, although the SMRC scheme has a significantly lower computational complexity when compared to the CMRC \cite{APS_Conceicao}, it can outperform the latter. Additionally, the PMRC scheme can also generally provide significantly better solutions in some cases, requiring much fewer RA loops, thus having a much lower time-complexity. However, in some other cases, the final solutions can be underestimated.

\subsection{Assessing the adaptability of the AP Selection Scheme}

In this section, we aim to evaluate the capacity of adaptability of the sequential AP selection scheme strategy when the RS network changes its configuration. For this, we perform the GA optimization for two different instances of network change. 

\begin{itemize}
    \item Instance 1: A new UE is added in a random position in the RS network, which increases the total number of UEs to $K{+}1$. For each scenario defined in the previous section, a newly added UE is placed in randomly selected coordinates. Their position vectors in the 3D space are $\left[6.67,0.5,1\right]$ (refer to Fig. \ref{fig:APS_scenario1}), $\left[13.2,19.5,1\right]$ (refer to Fig. \ref{fig:APS_scenario2}), $\left[9.55,11.8,1\right]$ (refer to Fig. \ref{fig:APS_scenario3}), and $\left[8.15,15.8,1\right]$ (refer to Fig. \ref{fig:APS_scenario4}) for scenarios 1, 2, 3, and 4, respectively.
    \item Instance 2: A random UE is removed from the RS network, which decreases the total number of UEs to $K{-}1$. In this case, for each scenario, a randomly selected UE is removed. The removed UEs were the ones placed in coordinates $\left[5.14,19.4,1\right]$ (refer to Fig. \ref{fig:APS_scenario1}), $\left[14.3,12.5,1\right]$ (refer to Fig. \ref{fig:APS_scenario2}), $\left[19.5,17.6,1\right]$ (refer to Fig. \ref{fig:APS_scenario3}), and $\left[16.6,15,1\right]$ (refer to Fig. \ref{fig:APS_scenario4}) for scenarios 1, 2, 3, and 4, respectively. 
\end{itemize}

In order to measure the adaptability of the aforementioned approach for the two new network instances, we take advantage of the information regarding the AP-UE association matrix that was optimized in the original scenario where we had $K$ UEs. To do this, we simply add a new row of zeros to this matrix and we take this solution as the initial one. We also consider the case where we do not retain any information from the original scenario and we simply rerun the GA optimization for the two new network instances. Therefore, we compare the cases where the GA has two different initial solutions: one where they are randomly selected and one where we include the original AP-UE association matrix. 

\begin{figure}[t!]
\centering
\subfigure[]{\includegraphics[width=1\columnwidth]{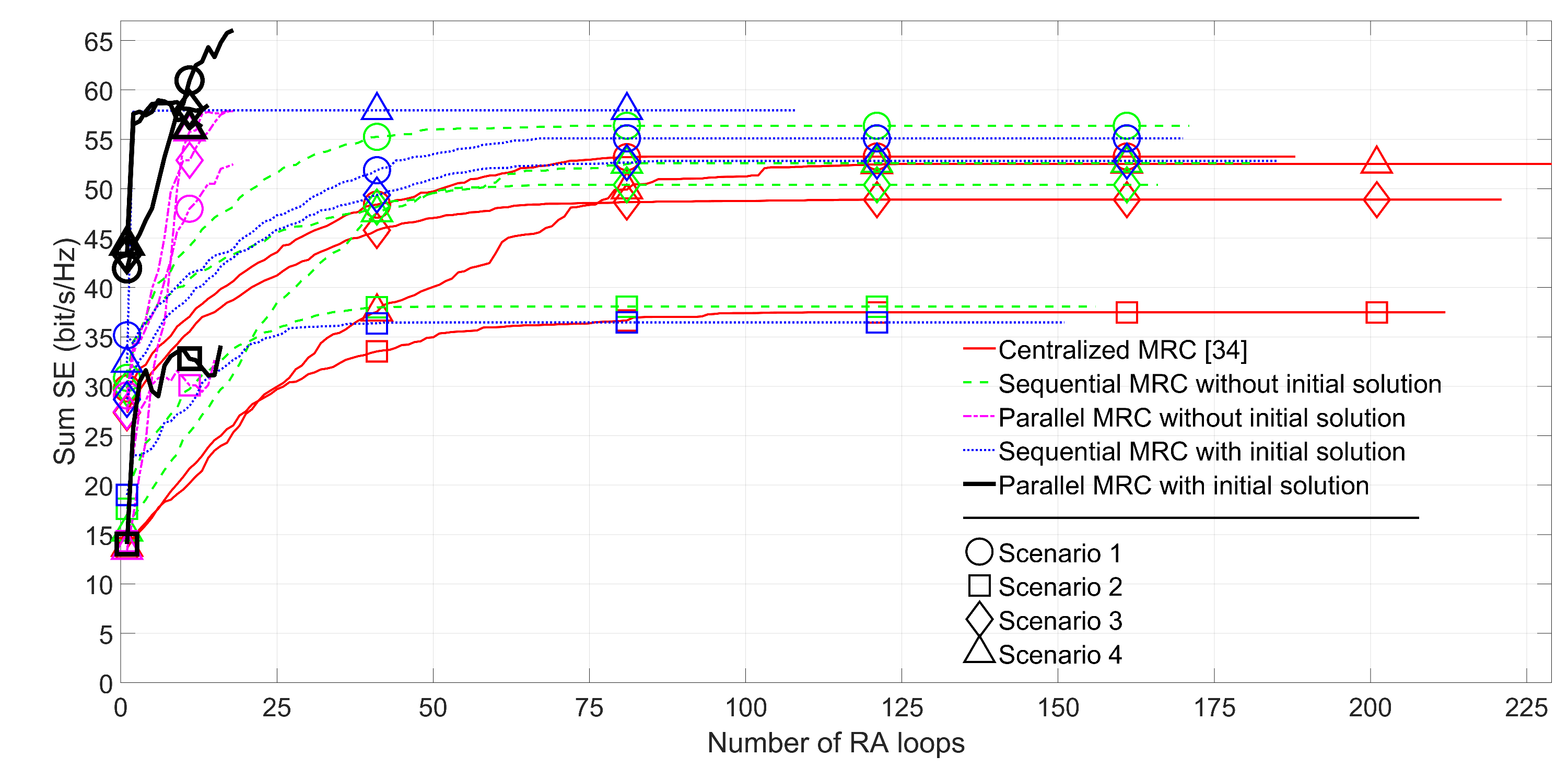}}
\hfill
\subfigure[]{\includegraphics[width=1\columnwidth]{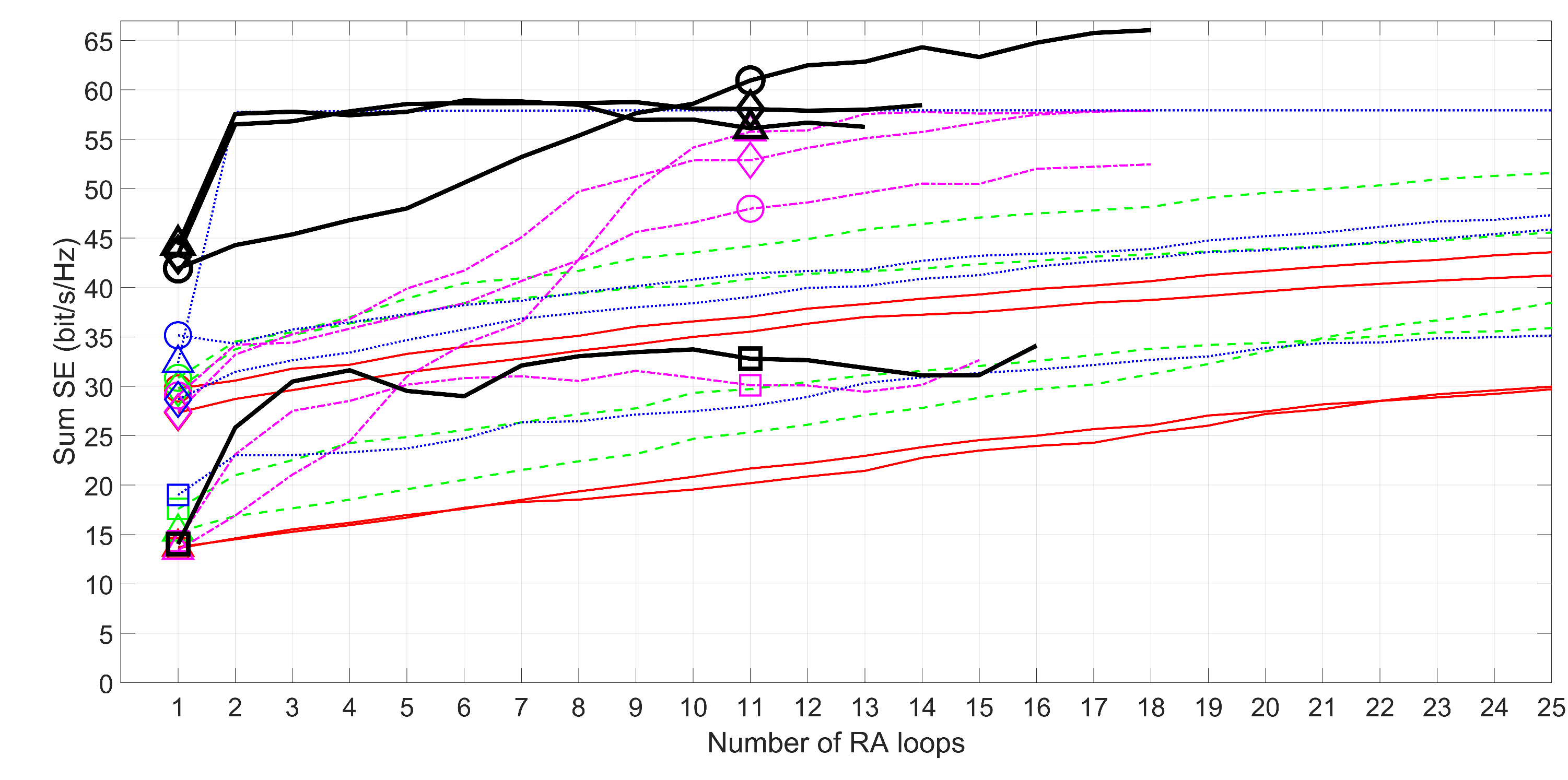}}
\caption{Sum SE, in bits/s/Hz, obtained with CMRC \cite{APS_Conceicao}, SMRC, and PMRC with and without an initial solution for instance 1: (a) full range and (b) zoomed in the 0-25 RA range.}
\label{fig:evolution_add}
\end{figure}

Fig. \ref{fig:evolution_add} illustrates the performance of the GA-based AP selection algorithm with the CMRC, SMRC, and PMRC with and without the initial solution for instance 1. Once again, we plot the sum SE (in bits/s/Hz) averaged through the $N_a$ realizations as a function of the number of RA loops. 

For scenario 1, at $10$ RA loops, the CMRC reached a sum SE of $36.6$ bits/s/Hz while the SMRC with and without an initial solution has a sum SE of $40.8$ and $43.5$ bits/s/Hz, respectively. The equivalent values for the PMRC with and without an initial solution are $58.6$ and $46.6$ bits/s/Hz, respectively. The CMRC converges to $53.3$ bits/s/Hz at around $86$ RA loops, while the SMRC with and without an initial solution converges to $55.1$ and $56.4$ bits/s/Hz around $71$ and $72$ RA loops. The PMRC with and without an initial solution converges to $66.0$ and $52.5$ bits/s/Hz both at $18$ RA loops.

For scenario 2, at $10$ RA loops, the CMRC reached is able to obtain a sum SE of $20.8$ bits/s/Hz while the SMRC with an initial solution has a sum SE of $27.5$ bits/s/Hz. The one without an initial solution has $29.3$ bits/s/Hz at this stage. The PMRC with and without an initial solution presents a sum SE of $33.7$ and $30.9$ bits/s/Hz, respectively. The CMRC converges to $38.1$ bits/s/Hz at around $57$ RA loops, while the SMRC with and without an initial solution converges to a final $36.5$ bits/s/Hz and $38.1$ bits/s/Hz around $52$ and $57$ RA loops. The PMRC with and without an initial solution once again converges much faster to $34.1$ and $32.6$ bits/s/Hz, requiring only $16$ and $15$ RA loops.

For scenario 3, at $10$ RA loops, the CMRC reaches a sum SE of $35.0$ bits/s/Hz while the SMRC with and without an initial solution has a sum SE of $38.4$ and $40.1$ bits/s/Hz, respectively. At this stage, a similar analysis to PMRC leads to $58.1$ and $52.9$ bits/s/Hz, respectively. The CMRC converges to $48.9$ bits/s/Hz at around $119$ RA loops, while the SMRC with and without an initial solution both converge to a final $52.8$ bits/s/Hz at around $86$ RA loops, respectively. The PMRC with and without an initial solution converges to $58.5$ and $57.9$ bits/s/Hz at around $14$ and $18$ RA loops.

For scenario 4, the SMRC with an initial solution converges almost instantly at $9$ RA loops, displaying a sum SE of $57.9$ bits/s/Hz. At $10$ RA loops, the CMRC shows a sum SE of $19.6$ bits/s/Hz while the SMRC without an initial solution has $24.7$ bits/s/Hz. The PMRC with and without an initial solution at this stage exhibits $57.0$ and $54.2$ bits/s/Hz, respectively. The CMRC converges to $52.5$ bits/s/Hz at around $127$ RA loops, while the SMRC without an initial solution converges to a final $52.6$ bits/s/Hz around $82$ RA loops and the PMRC with and without the solution has a final SE of $56.3$ and $57.9$ bits/s/Hz at $13$ and $18$ RA loops.

We can conclude that the SMRC with an initial solution is slightly effective for scenario 2 and greatly effective for scenario 4, which are the ones with $16$ antennas. This means that, in these scenarios, the optimized configuration of the AP-UE association matrix is almost the same as the one obtained for the original instance. In general, the fact that we add a random UE in the RS does not force the other APs to adjust their allocation vector as they remain similar to the ones in the original AP-UE association matrix. The same cannot be said for scenarios 1 and 3, where the number of total antennas is $48$. The newly added UE results in a relatively different configuration of the AP-UE association matrix, which significantly hinders the effect of employing the one obtained in the original instance as an initial solution. In these cases, there is no great benefit of using the SMRC with an initial solution as slightly better performances can be obtained with it when compared to SMRC without an initial solution. The behaviour of the PMRC approach either with or without an initial solution is irregular. This can be explained by the fact that in this approach, the calculated sum SE is not directly equal to the actual sum SE of the RS network, as explained earlier. This sometimes leads to AP-UE RA configurations which provide an efficient sum SE when the APs work independently, but when we consider the actual sum SE of the combined network, this solution is not ideal. However, the same cannot be said for PMRC, as we can observe that the PMRC option with an initial solution is effective for all scenarios, converging almost instantly while showing the best value of sum SE, when compared to all other approaches.

\begin{figure}[t!]
\centering
\subfigure[]{\includegraphics[width=1\columnwidth]{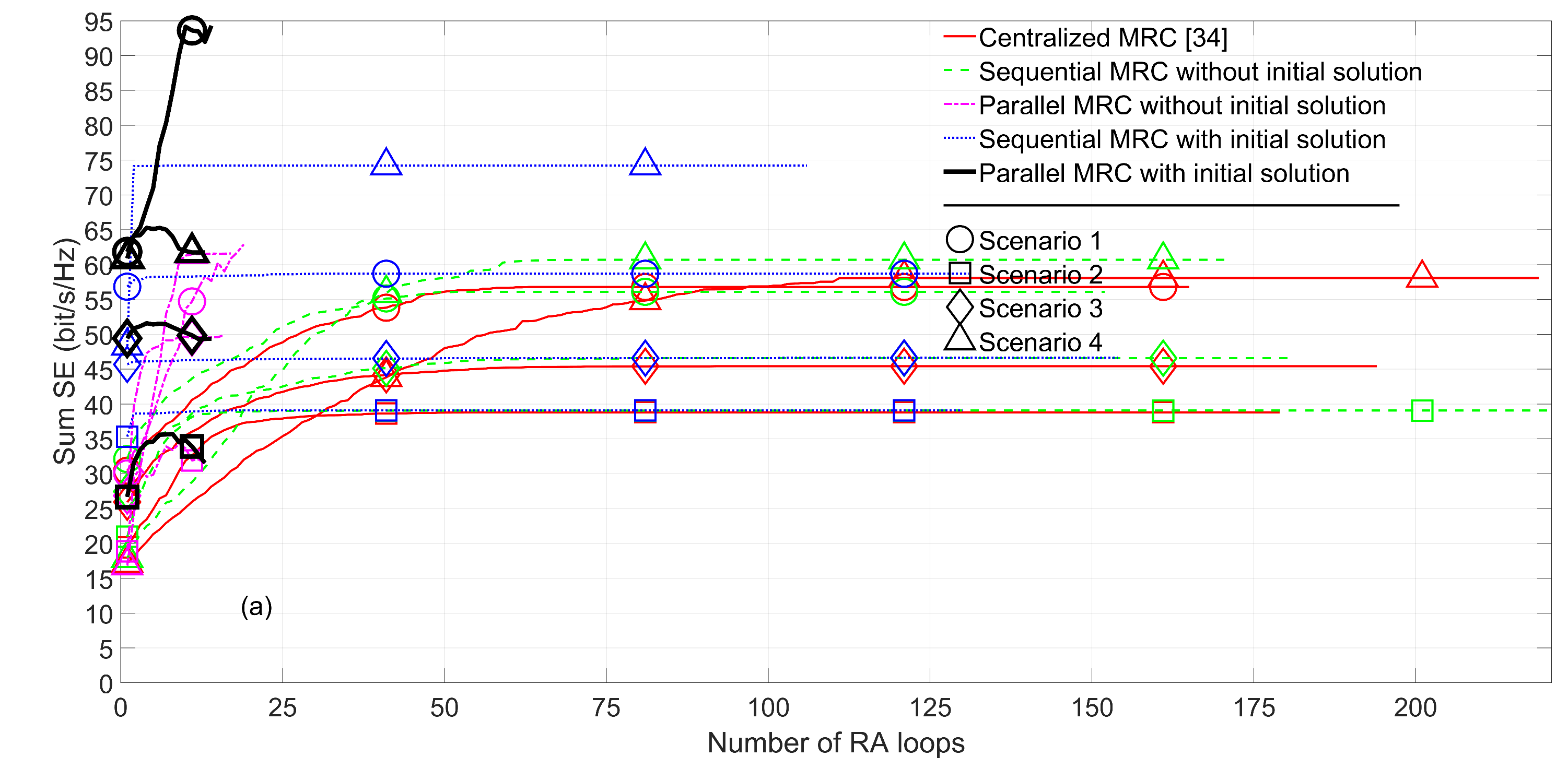}} 
\hfill
\subfigure[]{\includegraphics[width=1\columnwidth]{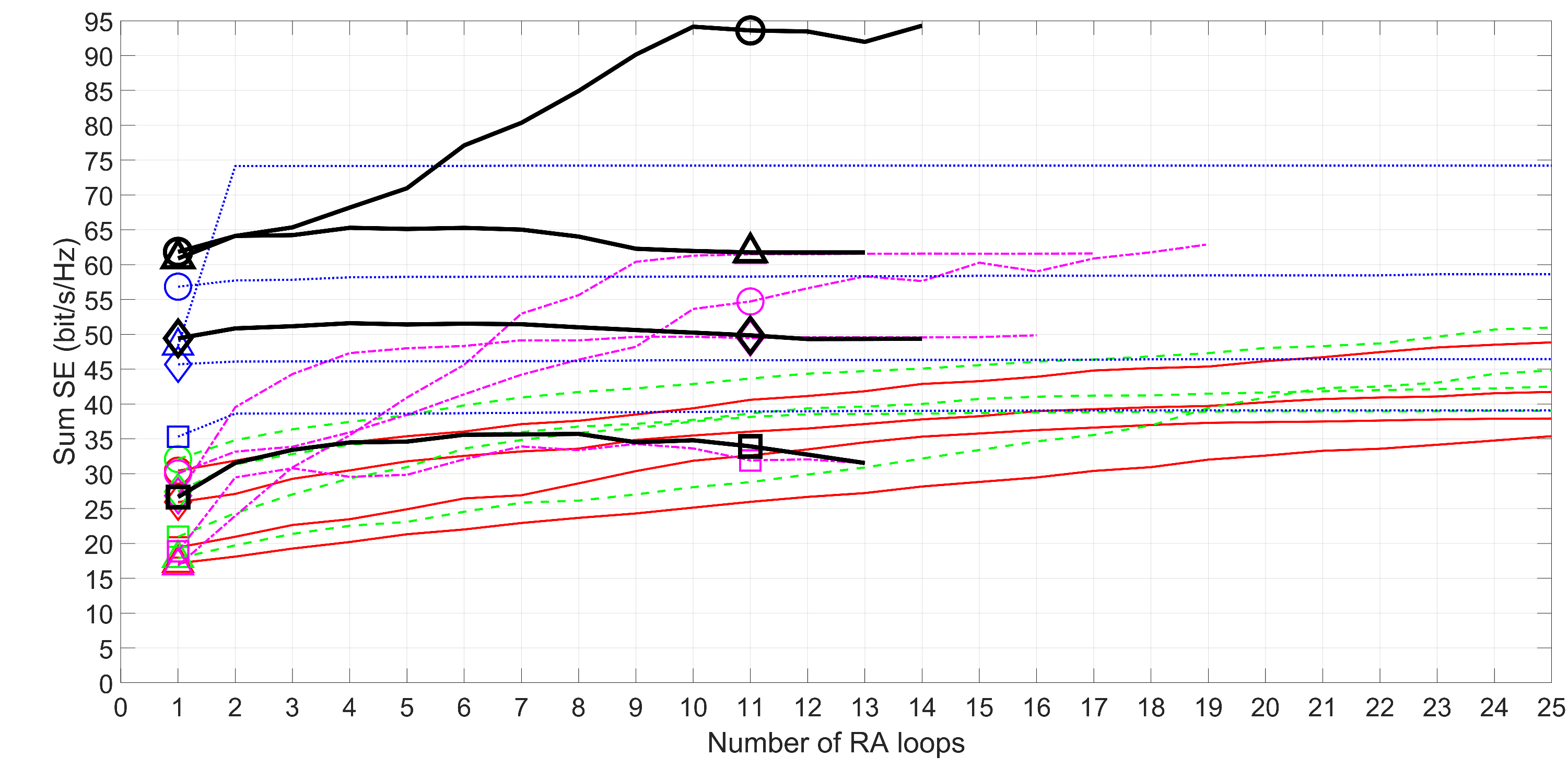}}
\caption{Sum SE, in bits/s/Hz, obtained with CMRC \cite{APS_Conceicao}, SMRC, and PMRC with and without an initial solution for instance 2: (a) full range and (b) zoomed in the 0-25 RA range.}
\label{fig:evolution_remove}
\end{figure}

Finally, the same analysis is done for instance 2 in Fig. \ref{fig:evolution_remove}. 

For scenario 1, at $10$ RA loops, the CMRC is only able to obtain a sum SE of $39.4$ bits/s/Hz while the SMRC with and without an initial solution has $58.3$ and $42.9$ bits/s/Hz. The same values for PMRC are $94.1$ and $53.6$ bits/s/Hz respectively. The CMRC converges to $56.8$ bits/s/Hz at around $65$ RA loops, the SMRC with and without an initial solution converges to $58.7$ and $56.1$ bits/s/Hz around $31$ and $52$ RA loops, while the PMRC with and without an initial solution converge to $94.3$ and $62.9$ bits/s/Hz around $14$ and $19$ RA loops. 

For scenario 2, at $10$ RA loops, the sum SEs of CMRC, SMRC, and PMRC with and without an initial solution are $31.9$, $38.9$, $37.5$, $34.8$, and $33.6$ bits/s/Hz, respectively. They converge to $38.8$, $39.1$, $39.1$, $31.5$ and $31.5$ bits/s/Hz at around $79$, $30$, $121$, $13$ and $13$ RA loops, respectively.

For scenario 3, at $10$ RA loops, the sum SEs of CMRC, SMRC, and PMRC with and without an initial solution are $35.5$, $46.3$, $37.7$, $50.2$, and $37.7$ bits/s/Hz, respectively. They converge to $45.4$, $46.6$, $46.6$, $49.4$ and $49.9$ bits/s/Hz at around $94$, $54$, $81$, $14$ and $16$ RA loops, respectively. 

For scenario 4, at $10$ RA loops, the sum SEs of CMRC, SMRC, and PMRC with and without an initial solution are $25.1$, $74.2$, $28.1$, $62.0$, and $61.3$ bits/s/Hz, respectively. They converge to $58.1$, $74.2$, $60.7$, $61.8$ and $61.6$ bits/s/Hz at around $119$, $6$, $71$, $13$ and $17$ RA loops, respectively. 

Unlike the previous instance, we can conclude that the SMRC with an initial solution is suitable to be employed for all scenarios. The reason behind this claim takes into account the similarities between the optimized configurations of the AP-UE association matrices before and after a UE is removed, as in general, removing a specific UE does not affect the performance of the remaining UEs. In other words, already established allocations for the other UEs are a good approximation of the optimal AP-UE configuration. The same can be applied to PMRC with an initial solution. Nonetheless, for scenarios with a high number of antennas, i.e. scenarios 1 and 3, this approach was able to outperform the SMRC with an initial solution as, in these cases, the total number of possible AP-UE allocation solutions in the PMRC method is much lower when compared to the SMRC case. For scenarios 2 and 4, due to the relatively lower number of total possible solutions, the SMRC with an initial solution is more efficient due to the fact that the PMRC method ignores the inter-AP interference in signal detection.

\section{Conclusions}

In this work, we have proposed sequential and a parallel MRC-based RA strategy for an RS network that focuses on choosing the best allocation between each UE and APs' antenna to maximize the total SE. We have compared their performances to a centralized approach with two different equalization schemes: CMRC and COSLP. To solve the optimization problem and find an effective solution for the AP-UE association matrix, we have proposed a low-complexity MH algorithm based on a GA. 

The COSLP is the one that provides the best SE despite exhibiting the highest time-complexity and fronthaul signalling capacity. The CMRC can slightly reduce the total number of RA loops, improving both metrics. However, since it is based on the MRC scheme, the SE gains are lower. The SMRC lowers the computational complexity and fronthaul signalling from the CMRC approach, presenting faster convergence to the final solution. Finally, PMRC requires the lowest fronthaul signalling capacity among all compared approaches. It can also significantly improve the sum SE and convergence speed compared to all other methods. However, the parallel nature can hinder its performance in some scenarios, leading to inaccurate RA solutions.

In low-density RS networks, we have concluded that the SMRC scheme can leverage an initial UE-AP configuration to boost the performance of the AP selection optimization for instances when new UEs want to connect to the RS. The PMRC does not depend on this density and efficiently leverages an initial solution to improve the convergence speed of the AP selection scheme. Additionally, regardless of the scenario, the effectiveness of combining the SMRC with the initial solution is also shown for instances when some UEs may leave the RS network. The PMRC can also outperform the SMRC, but only for high-density scenarios where the search space of solutions is much smaller.

Our results indicate that the proposed algorithm demonstrates strong performance within small time-frames, making it particularly effective in scenarios with limited time sensitivity. Furthermore, our findings suggest that minor variations in the network scenario (such as introducing a new UE or removing a current UE from the network) do not substantially impact the solutions provided by the algorithm, as the original AP selection solution from the original scenario is beneficial in most cases. This resilience to small changes means that the algorithm remains practical and efficient in many real-world applications, even when some degree of environmental dynamism is present. Therefore, although the algorithm may need to be rerun when larger environmental shifts occur, its ability to maintain performance despite minor fluctuations adds a layer of practicality that addresses concerns about time sensitivity.

\begin{IEEEbiography}[{\includegraphics[width=1in,height=1.25in,clip,keepaspectratio]{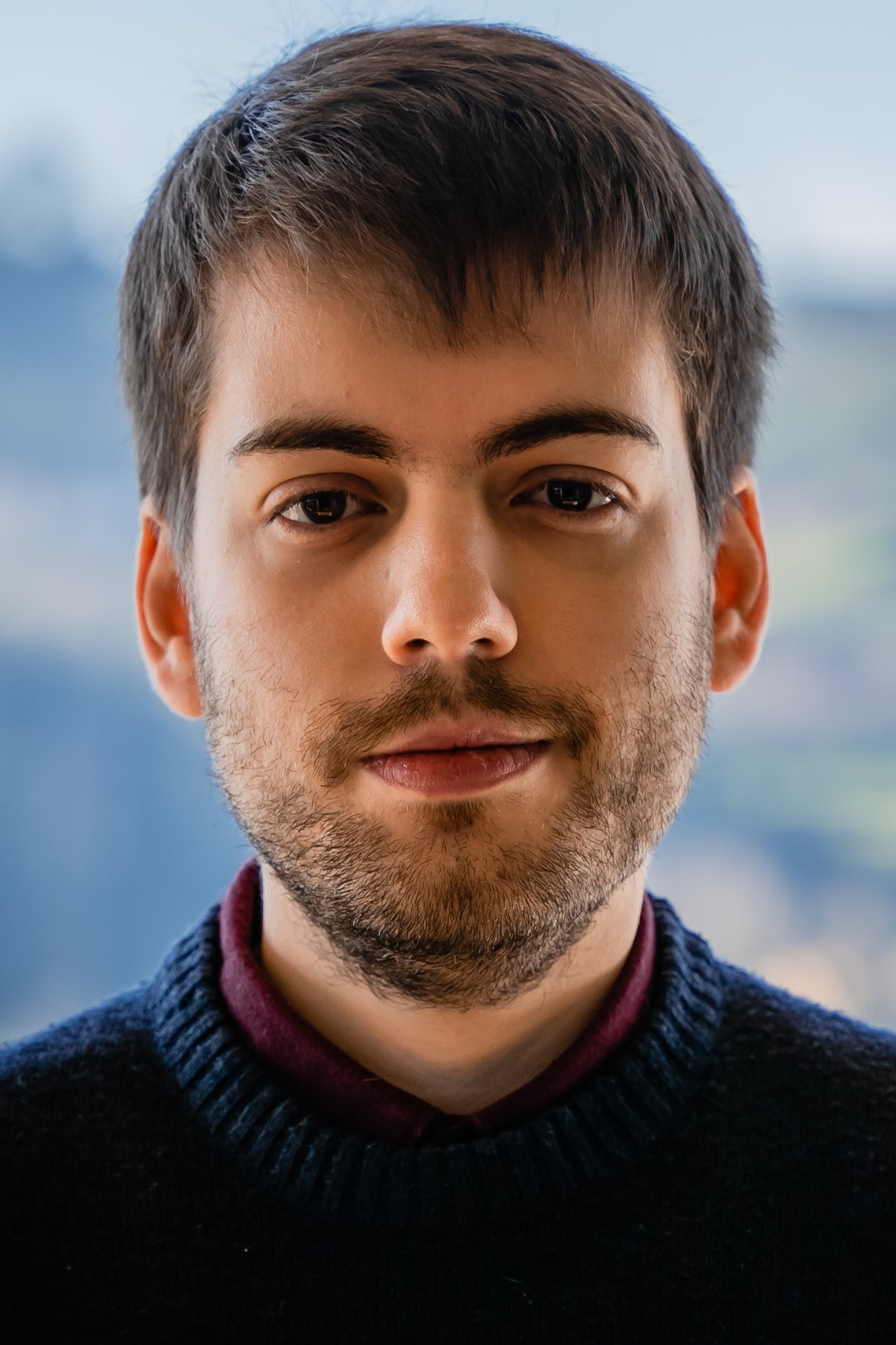}}]{FILIPE CONCEIÇÃO } (Graduate Student Member, IEEE) was born in Coimbra, Portugal, in 1996. He received the M.Sc. degree in electrical and computer engineering specialty in telecommunications from the University of Coimbra, in 2019. Currently, he is pursuing a Ph.D. degree at the Department of Electrical and Computer Engineering, Faculty of Science and Technology from the University of Coimbra and Instituto de Telecomunicações (IT) focusing on the development of transmission, detection, and resource allocation techniques for radio stripe systems. His current research interests include cooperative communications beyond 5G networks, including distributed massive MIMO, cell-free, radio stripe, resource allocation, and pre-coding/equalization techniques.   
\end{IEEEbiography}

\begin{IEEEbiography}[{\includegraphics[width=1in,height=1.25in,clip,keepaspectratio]{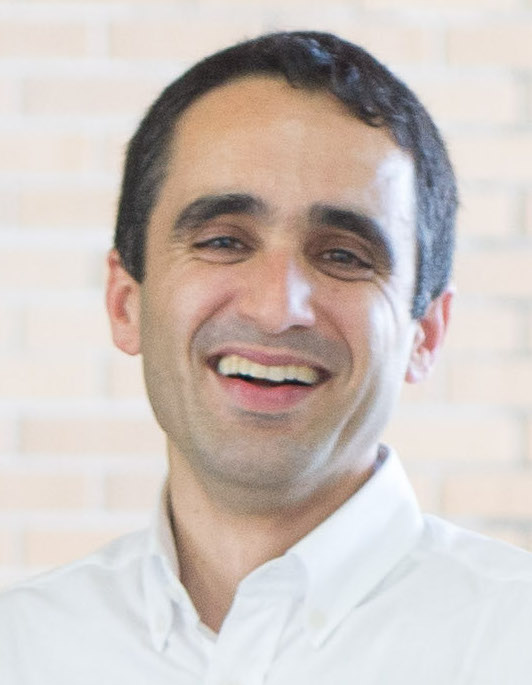}}]{MARCO GOMES } (Senior Member, IEEE) was born in Coimbra, Portugal, in 1977. He received the M.Sc. and Ph.D. degrees in electrical and computer engineering specialty in telecommunications from the University of Coimbra, in 2004 and 2011, respectively, where he is currently an Assistant Professor with the Department of Electrical and Computer Engineering, Faculty of Science and Technology. He is also a Senior Researcher with the Instituto de Telecomunicações (IT) and performs different collaborations under his research and development activities with other institutions and universities worldwide. His main research interests include wireless digital communications, massive-MIMO, large intelligent surfaces, general signal processing for communications, software-defined radio, error control coding and physical layer security. He is a member of the IEEE Communications Society and the IEEE Vehicular Technology Society. Dr. Gomes is also an Editor of the IEEE Communications Letters, the IEEE Open Journal of the Communications Society, the IEEE ComSoc On-Line Content, and Physical Communication (Elsevier).
\end{IEEEbiography}

\begin{IEEEbiography}[{\includegraphics[width=1in,height=1.25in,clip,keepaspectratio]{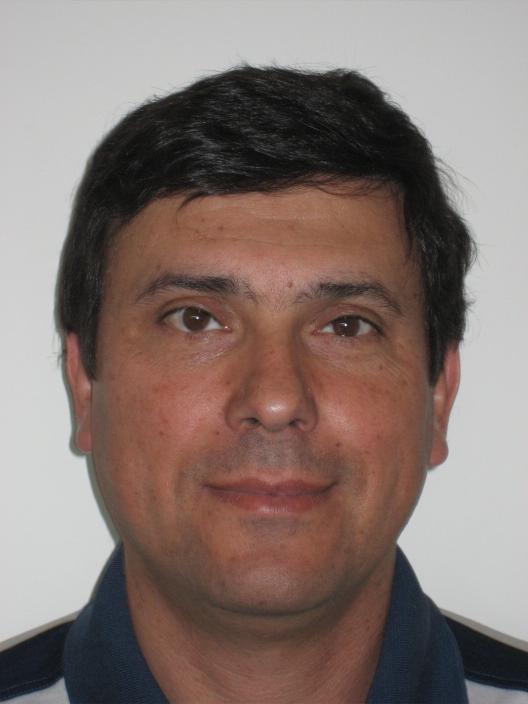}}]{VITOR SILVA } received the Graduation Diploma in electrical engineering and the Ph.D.degree from the University of Coimbra, Portugal, in 1984 and 1996, respectively. He is currently an Assistant Professor with the Department of Electrical and Computer Engineering, University of Coimbra, where he lectures signal processing and information and coding theory. He published more than 200 papers. His research activities include signal processing, parallel computing, and coding theory.
\end{IEEEbiography}

\begin{IEEEbiography}[{\includegraphics[width=1in,height=1.25in,clip,keepaspectratio]{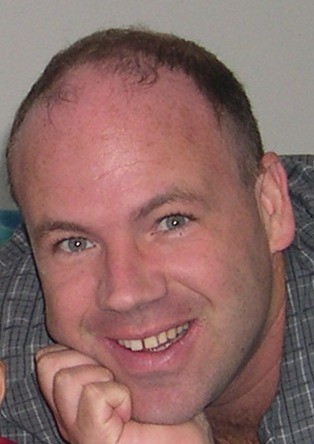}}]{RUI DINIS } (Senior Member, IEEE) received the Ph.D. degree from Instituto Superior Técnico (IST), Technical University of Lisbon, Portugal, in 2001 and the Habilitation in Telecommunications from Faculdade de Ciências e Tecnologia (FCT), Universidade Nova de Lisboa (UNL), in 2010. From 2001 to 2008 he was a Professor at IST. Currently, he is an associate professor at FCT-UNL. During 2003 he was an invited professor at Carleton University, Ottawa, Canada. He was a researcher at CAPS (Centro de Análise e Processamento de Sinal), IST, from 1992 to 2005 and a researcher at ISR (Instituto de Sistemas e Robótica) from 2005 to 2008. Since 2009 he is a researcher at IT (Instituto de Telecomunicações). He has been actively involved in several national and international research projects in the broadband wireless communications area. His research interests include transmission, estimation, and detection techniques. Rui Dinis is a VTS Distinguished Lecturer and is or was an editor at IEEE Transactions on Wireless Communications, IEEE Transactions on Communications, IEEE Transactions on Vehicular Technology, IEEE Open Journal on Communications, and Elsevier Physical Communication. He was also a guest editor for several special issues. 
\end{IEEEbiography}

\end{document}